\documentclass[astrosymb, twocolumn]{aastex631}

\usepackage{url}
\usepackage{graphicx}

\shorttitle{DR17 Brackett Emitters Catalog}
\shortauthors{Campbell, Khilfeh, et. al}

\graphicspath{{./}{figures/}}

\begin{document}

\title{Pre-main Sequence Brackett Emitters in the APOGEE DR17 Catalog: \\ Line Strengths and Physical Properties of Accretion Columns}

\author[0000-0001-5436-5388]{Hunter Campbell}
\affiliation{Department of Physics and Astronomy, Western Washington University, 516 High St, Bellingham, WA 98225, USA}

\author[0000-0001-9649-6028]{Elliott Khilfeh}
\affiliation{Department of Physics and Astronomy, Western Washington University, 516 High St, Bellingham, WA 98225, USA}

\author[0000-0001-6914-7797]{Kevin R. Covey}
\affiliation{Department of Physics and Astronomy, Western Washington University, 516 High St, Bellingham, WA 98225, USA}

\author[0000-0002-5365-1267]{Marina Kounkel}
\affil{Department of Physics and Astronomy, Vanderbilt University, VU Station 1807, Nashville, TN 37235, USA}
\affiliation{Department of Physics and Astronomy, Western Washington University, 516 High St, Bellingham, WA 98225, USA}

\author{Richard Ballantyne}
\affiliation{Department of Physics and Astronomy, Western Washington University, 516 High St, Bellingham, WA 98225, USA}

\author{Sabrina Corey}
\affiliation{Department of Physics and Astronomy, Western Washington University, 516 High St, Bellingham, WA 98225, USA}


\author[0000-0001-8600-4798]{Carlos G. Rom\'an-Z\'u\~niga}
\affiliation{Universidad Nacional Aut\'onoma de M\'exico, Instituto de Astronom\'ia, AP 106,  Ensenada 22800, BC, México}

\author[0000-0001-9797-5661]{Jes\'us Hern\'andez}
\affiliation{Universidad Nacional Aut\'onoma de M\'exico, Instituto de Astronom\'ia, AP 106,  Ensenada 22800, BC, México}

\author[0000-0001-6647-862X]{Ezequiel Manzo Martínez}
\affiliation{Universidad Nacional Aut\'onoma de M\'exico, Instituto de Astronom\'ia, AP 106,  Ensenada 22800, BC, México}

\author[0000-0002-5855-401X]{Karla Pe\~na Ramírez}
\affiliation{Centro de Astronomía (CITEVA), Universidad de Antofagasta, Av. Angamos 601, Antofagasta, Chile}

\author[0000-0002-1379-4204]{Alexandre Roman-Lopes}
\affiliation{Departamento de Astronomia, Facultad de Ciencias, Universidad de La Serena.  Av. Juan Cisternas 1200, La Serena, Chile}

\author[0000-0002-3481-9052]{Keivan G.\ Stassun}
\affil{Department of Physics and Astronomy, Vanderbilt University, VU Station 1807, Nashville, TN 37235, USA}

\author[0000-0003-1479-3059]{Guy S. Stringfellow}
\affiliation{Center for Astrophysics and Space Astronomy, Department of Astrophysical and Planetary Sciences, University of Colorado, Boulder,CO, 80309, USA}


\author[0000-0002-5936-7718]{Jura Borissova}
\affil{Instituto de Física y Astronomía, Universidad de Valparaíso, Av. Gran Bretaña 1111, Playa Ancha, Casilla 5030, Chile}
\affil{Millennium Institute of Astrophysics, Nuncio Monsenor Sotero Sanz 100, Of. 104, Providencia, Santiago, Chile}

\author[0000-0001-9984-0891]{S. Drew Chojnowski}
\affil{Department of Astronomy, New Mexico State University, Las Cruces, NM 88001, USA}


\author{Valeria Ram\'irez-Preciado}
\affiliation{Universidad Nacional Autónoma de México, Instituto de Astronom\'ia, AP 106,  Ensenada 22800, BC, México}

\author[0000-0001-6072-9344]{Jinyoung Serena Kim}
\affiliation{Steward Observatory, Department of Astronomy, University of Arizona, 933 North Cherry Avenue, Tucson, AZ 85721, USA}

\author[0000-0001-7351-6540]{Javier Serna}
\affiliation{Universidad Nacional Aut\'onoma de M\'exico, Instituto de Astronom\'ia, AP 106,  Ensenada 22800, BC, Mexico}

\author[0000-0003-2300-8200]{Amelia M.\ Stutz}
\affiliation{Departamento de Astronom\'{i}a, Universidad de Concepci\'{o}n,Casilla 160-C, Concepci\'{o}n, Chile}
\affiliation{Max-Planck-Institute for Astronomy, K\"{o}nigstuhl 17, 69117 Heidelberg, Germany}

\author[0000-0002-7795-0018]{Ricardo López-Valdivia}
\affiliation{Universidad Nacional Aut\'onoma de M\'exico, Instituto de Astronom\'ia, AP 106,  Ensenada 22800, BC, México}

\author[0000-0002-2011-4924]{Genaro Su\'arez}
\affiliation{Department of Physics and Astronomy, The University of Western Ontario, 1151 Richmond St, London, ON N6G 1N9, Canada}


\author[0000-0002-3576-4508]{Jason E. Ybarra}
\affiliation{Department of Physics and Astronomy, 
White Hall, Box 6315, West Virginia University, 
Morgantown, WV 26506-6315}


\author[0000-0001-9330-5003]{Pen\'elope Longa-Pe\~na}
\affiliation{Unidad de Astronom\'ia, Universidad de Antofagasta, Avenida Angamos 601, Antofagasta 1270300, Chile}

\author[00000-0003-3526-5052]{Jos\'e G. Fern\'andez-Trincado}
\affiliation{Instituto de Astronom\'ia, Universidad Cat\'olica del Norte, Av. Angamos 0610, Antofagasta, Chile}



\begin{abstract}
Very young (t $\lesssim$ 10 Myrs) stars possess strong magnetic fields that channel ionized gas from the interiors of their circumstellar discs to the surface of the star. Upon impacting the stellar surface, the shocked gas recombines and emits hydrogen spectral lines. To characterize the density and temperature of the gas within these accretion streams, we measure equivalent widths of Brackett (Br) 11 – 20 emission lines detected in 1101 APOGEE spectra of 326 likely pre-main sequence accretors. For sources with multiple observations, we measure median epoch-to-epoch line strength variations of 10\% in Br11 and 20\% in Br20. We also fit the measured line ratios to predictions of radiative transfer models by Kwan \& Fischer.  We find characteristic best-fit electron densities of $n_e$ = 10$^{11} - 10^{12}$ cm$^{-3}$, and excitation temperatures that are inversely correlated with electron density (from T$\sim$5000 K for $n_e \sim 10^{12}$ cm$^{-3}$, to T$\sim$12500 K at $n_e \sim 10^{11}$ cm$^{-3}$). These physical parameters are in good agreement with predictions from modelling of accretion streams that account for the hydrodynamics and radiative transfer within the accretion stream. We also present a supplementary catalog of line measurements from 9733 spectra of 4255 Brackett emission line sources in the APOGEE DR17 dataset. 

\end{abstract}

\keywords{Stellar accretion (1578) --- Emission line stars (460) --- Protostars (1302) --- Young stellar objects (1834) --- Stellar spectral lines (1630)}

\section{Introduction} \label{sec:intro}

Early on, young stellar objects (YSOs) have massive protoplanetary disks through which they accrete gas. The material flows through the disk and, upon reaching the disk's inner edge, travels along the magnetic field lines onto the stellar surface \citep{hartmann2016}. In the process, outflows are launched and remove excess angular momentum inherited from the rotating disk.

When this accreting material impacts the stellar photosphere, it forms a shock that emits both continuum and line emission from the heated and excited gas.   Immediately following a particularly strong accretion event, the YSO may increase in brightness by several orders of magnitude, producing an FU Ori-type outburst \citep{bae2014}. While typical accretion rates are significantly lower than in the one observed in these events to be sustainable over the lifetime of protoplanetary disks, even weak accretion signatures are nonetheless imprinted on the stellar spectra.

Most notable accretion signatures are reflected in form of hydrogen emission lines, particularly H$\alpha$. Classical T Tauri stars (CTTSs) have wide H$\alpha$ lines, with equivalent widths (EqWs) in excess of what is found in weak lined T Tauri stars (WTTSs) \citep{white2003}. This is due to CTTSs having active ongoing accretion and outflows, whereas in WTTSs the emission is driven solely by the magnetic activity on the photosphere. In these WTTSs, the accretion is terminated, and the protoplantary disks are largely depleted.

H$\alpha$ is a particularly useful line to observe accretion because it is the lowest energy transition of hydrogen that is easily observable, and it is easy to excite even with low accretion/outflow rates. However, other H lines, both higher level lines in the Balmer series, as well as in other series, can also often be observed in emission in particularly active accretors. Through using multiple transitions it is possible to determine the properties of the infalling gas, such as its temperature and density \citep{KwanFischer2011,Antoniucci2017,Gutierrez2020}. In turn, the bulk census of sources for which these parameters are known yields better constraints on the accretion models of YSOs.

In this paper, we use APOGEE spectra to identify sources with Brackett emission lines that fall in its spectral range in H band, 11 through 20. In Section \ref{sec:observations} we describe the data used for this analysis. In Section \ref{sec:measurements} we measure the equivalent widths of these lines. Then, in Section \ref{sec:models} we perform model fitting to these measurements to constrain the properties of the gas that is responsible for the emission. In Section \ref{sec:disc} we discuss the results, and in Section \ref{sec:conclusions} we offer the conclusions.

\section{Data and Methods}\label{sec:data}

\subsection{Observations} \label{sec:observations}


We analyze high-resolution (R $\sim$ 22,500) near-infrared (1.51–1.70 $\mu$m) spectra taken with the 300-fiber APOGEE spectrograph \citep{Wilson2019} on the Sloan 2.5 meter telescope \citep{Gunn2006} at Apache Point and Las Campanas Observatories. To generate a comprehensive, uniform and complete catalog of Brackett emission sources, we process the full contents of the SDSS-IV Data Release 17 \citep{Abdurro'uf22}. The APOGEE DR17 dataset includes all previously released APOGEE and APOGEE-2 spectra; in total, the dataset contains 2,659,178 individual spectra, all of which have been re-reduced with the latest version of the APOGEE data reduction and analysis pipeline \citep[][J. Holtzman et al. in prep.]{Nidever2015, Holtzman2015, GarciaPerez2016, Jonsson2020}.

Brackett emitters should be relatively rare in this dataset, however, as most stars targeted by APOGEE are red giants, whose luminosity makes them a superb tracer for dissecting the structure, dynamics, chemical evolution, and star formation history of the Milky Way \citep{Majewski2017}. The APOGEE-2 footprint is designed to sample these red giants along  systematic tiling of sightlines through the Galaxy's bulge, disk, and halo \citep{Blanton2017, Zasowski2013, Zasowski2017}, but Brackett emitters, particularly the bona fide accreting YSOs which are of most scientific interest in this paper, are concentrated in a small number of plates that overlap nearby star forming regions. Some standard APOGEE and APOGEE-2 plates may overlap with these star forming regions serendipitously, and can be identified in the survey tiling maps presented by \citet{Santana2021} and \citet{Beaton2021} in their summaries of the final targeting strategies for the southern and northern APOGEE programs, respectively. We expect the vast majority of the bona fide accreting YSOs to be identified outside of the standard survey tiling, however, on plates observed as part of the APOGEE-1 IN-SYNC ancillary program \citep{Cottaar2014, Foster2015,DaRio2016}, the APOGEE-2 Young Cluster Survey \citep[][ and Roman-Zuniga, in rev.]{Cottle2018, Kounkel2018}, or through several APOGEE-2 Contributed Programs targeting other star forming regions \citep[e.g., ][]{Borissova2019, RomanLopes20, Medina2021}. We tabulate in Table \ref{tab:yc_plates} the specific set of Young Cluster Plates that we focus our search and analysis efforts on. 

\startlongtable
\begin{deluxetable}{lrr}
\tabletypesize{\footnotesize}
\tablecolumns{12}
\tablewidth{0pt}

\tablecaption{Young Cluster Plates\label{tab:yc_plates}}
\tablehead{
  \colhead{Region} &
  \colhead{Plates} & 
  \colhead{$N_{\rm Br}$\tablenotemark{$^1$}}
}
\startdata
  IC348 & 6218-6223, 7073-7075, 7079, & 12\\
   &  10100, 12706 & \\
  NGC1333 & 6224-6226, 7070-7072, 11425 & 7\\
  NGC2264 & 6103, 6227, 10302, 11778 & 17\\  
  Orion-A & 7220-7234, 9481, & 98\\
  & 9533, 9659-9661, & \\ 
  & 11593-11594, 12273 & \\
  Orion-B & 8890-8899 & 19\\
  Orion-OB1ab & 8900-8906, 9468-9480 & 14\\
  Lambda Ori & 8879-8887, 9482, & 6\\
   & 9537, 9538 \\ 
  Pleiades & 8889, 9257 & 0\\
  Taurus & 9258-9259, 9287-9288, & 8\\
  & 11426-11431 & \\
  Cor-Aus & 10718 & 0\\
  Vela Ridge & 12276 & 7\\
  Rosette & 11440-11441, 12266 & 2\\
  h and $\chi$ Persei & 9244 & 5\\
  California Neb. & 11432-11438 & 6\\
  Carina & 9752, 10294-10298, & 56\\
         &  11620-11622, & \\
  & 12356-12358 & \\
  Cygnus X & 11271-11276, 11409-11420 & 69\\
\enddata
\tablenotetext{1}{Number of sources with detectable Br11 emission in a given region, excluding double-peaked or nebular sources}
\end{deluxetable}

\subsection{Line and Line Ratio Measurements} \label{sec:measurements}

\subsubsection{Spectral Preparation} 

To prepare the spectra for EqW and line ratio measurements, we clean them of cosmetic artifacts (i.e, chip gaps and sky line residuals) that can contaminate the line or continuum regions used in our measurements, and then place each spectrum in a consistent barycentric velocity frame. We do not attempt to correct to the rest frame of the pre-main sequence star, however, as the strong emission lines within the spectrum often degrade the quality of the radial velocity measurement of the individual visits produced by the APOGEE pipeline.

We use individual visit spectra in our analysis, to preserve information about variations in the strength or velocity profile of the Brackett lines. The spectra in the apVisit files are presented in the observed, geocentric velocity frame. We mask regions associated with prominent telluric emission lines to eliminate any residuals that may be left in the spectrum. For this process, we utilize the line list of telluric features in the APOGEE bandpass first used by \citet{Canas2018} to mask APOGEE spectra of Kepler 503-b prior to radial velocity measurement. 

APOGEE spectra are recorded using 3 distinct detectors, each covering a different range of wavelengths (red, green, blue). The cutoffs between the detectors occur near the Br12 and Br15 lines. Depending on the velocity of the source, these spectral gaps can fall within the continuum regions used to measure the Br12 and Br15 EqWs. We therefore interpolate over the wavelengths that fall in the gaps between detectors, using the mean values of the pixels on either edge. However, we independently estimate the signal-to-noise ratio (SNR) for each detector for each observation.

To determine the wavelength correction required to place these spectra in the barycentric frame, we used the pyAstronomy \texttt{helcorr} package\footnote{see \url{https://pyastronomy.readthedocs.io/en/latest/pyaslDoc/aslDoc/baryvel.html\#PyAstronomy.pyasl.helcorr} and \url{https://github.com/sczesla/PyAstronomy} } \citep{Czesla} to calculate the barycentric correction, based on the original IDL implementation by \citet{Piskunov2002}. This function uses the target coordinates and the time and location of observation. We adopt the following longitude, latitude, and elevations for each site: APO - long. $= -105.4913^{\circ}$, lat. $= 36.4649^{\circ}$, elevation $= 2788$m ; LCO - long. $= -70.413336^{\circ}$, lat. $= -29.05256^{\circ}$, elevation $= 2380$m. 

\begin{figure*}
\begin{center}
\includegraphics[scale=0.5]{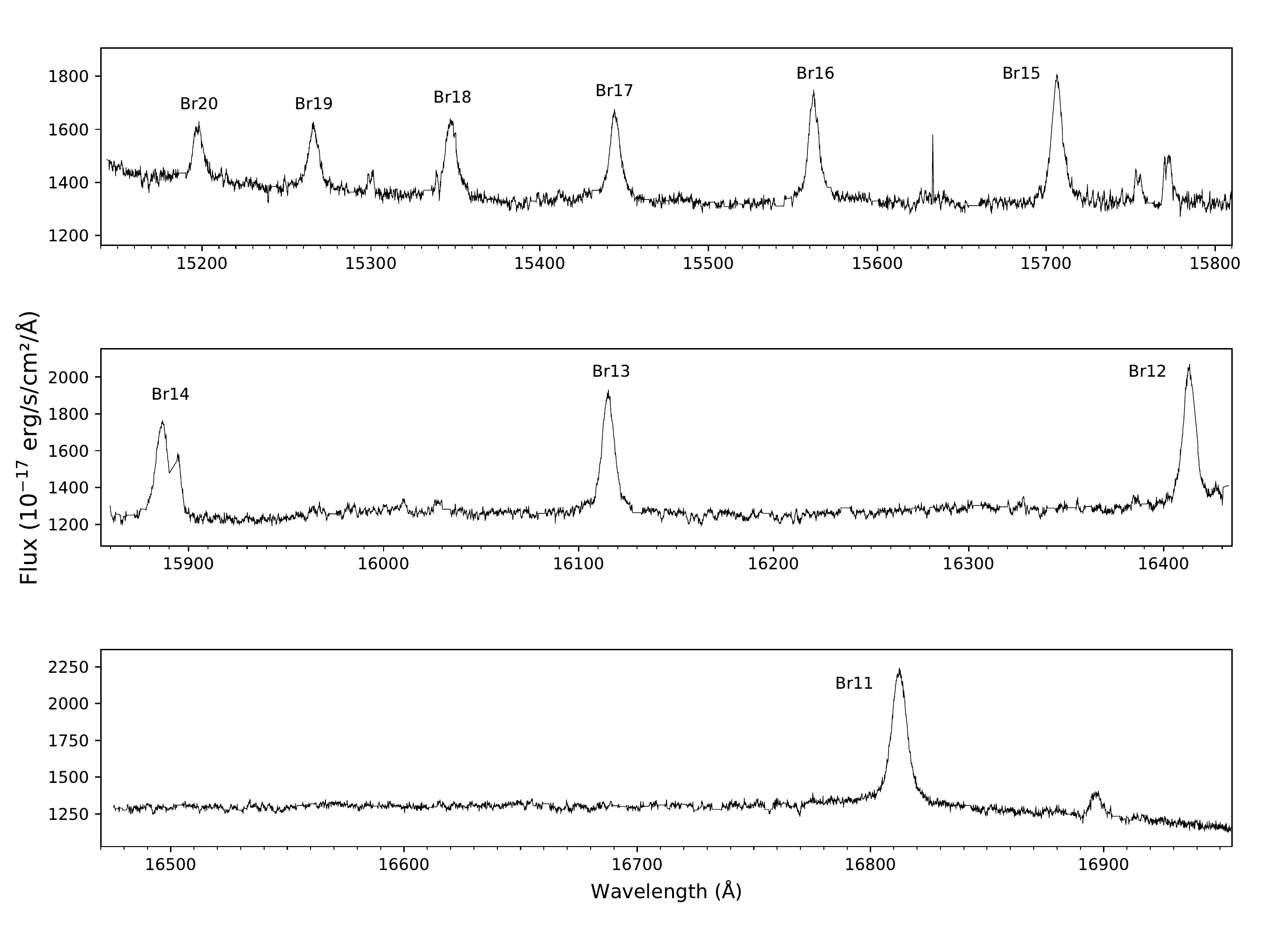}
\caption{Full spectrum plot of the Br11-20 emission lines for a strongly accreting YSO, 2M05471411+0009073. \label{fig:Full}}
\end{center}
\end{figure*}

Figure \ref{fig:Full} shows a final corrected spectrum for 2M05471411+0009073, a sample YSO with strong Brackett emission lines.

\subsubsection{Line Measurement}

\begin{figure*}
\begin{center}
\includegraphics[width=0.62\linewidth]{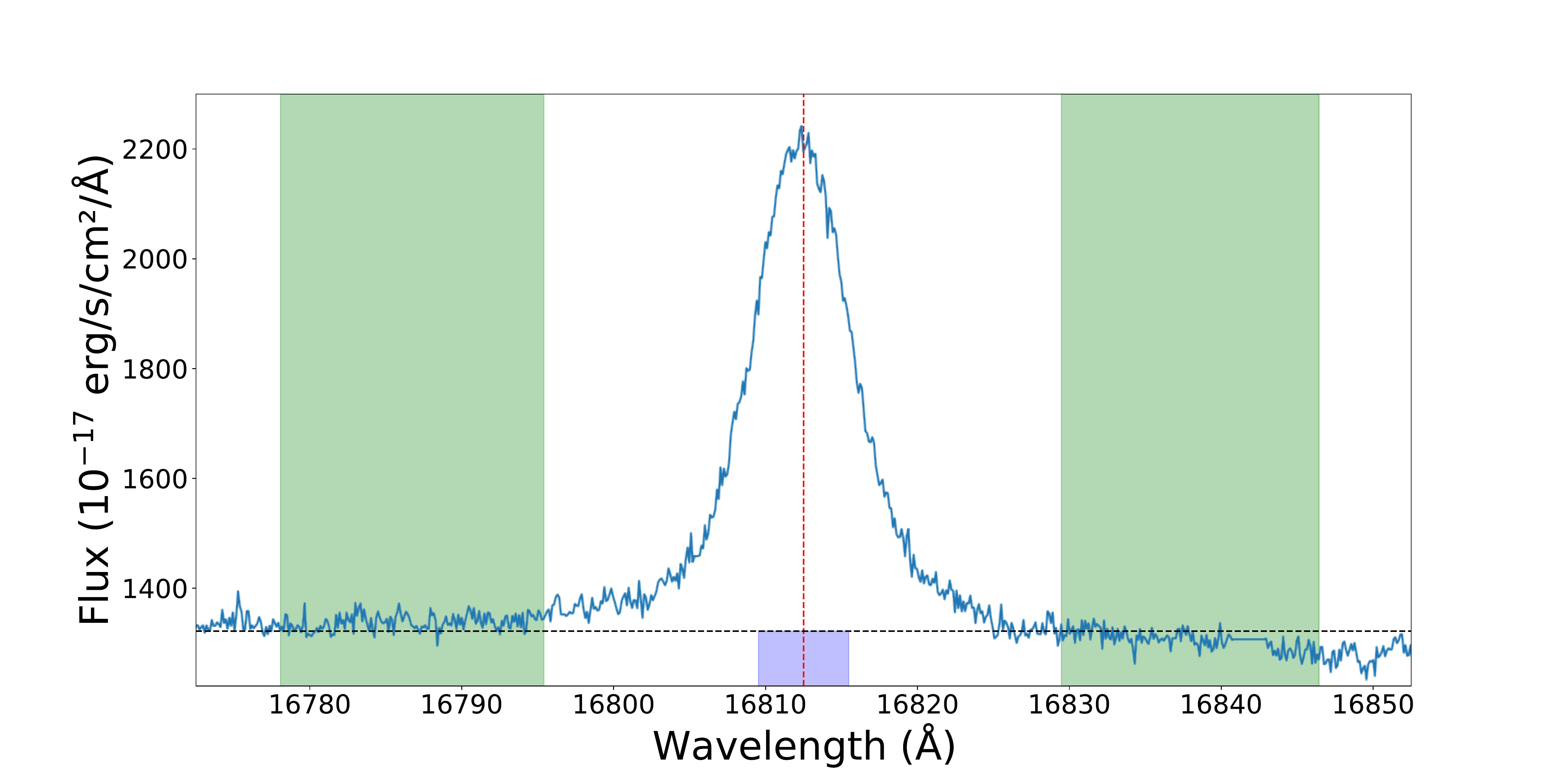}
\includegraphics[width=0.37\linewidth]{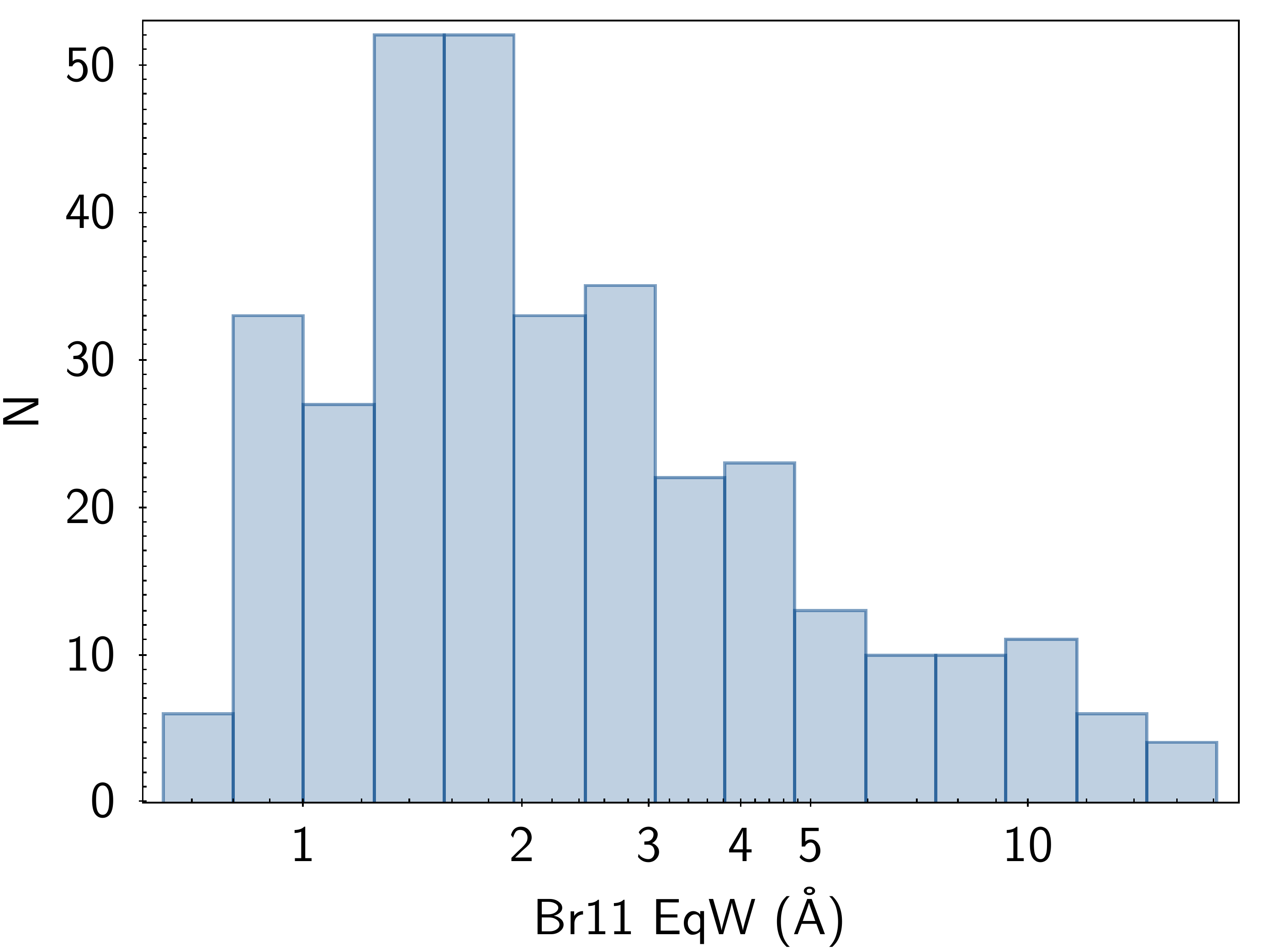}
\caption{Left: Wavelength-flux plot of the Br11 emission line for a strongly accreting YSO, 2M05471411+0009073. The green highlighted regions are the windows which were used to calculate the continuum (horizontal dashed line) on the left and right of the central line (vertical dashed line). The equivalent width of the Br11 line is given by the width of the shaded blue region below the continuum. Right: Distribution of the absolute value of equivalent widths for Br11 for all of the objects in the sample, excluding those that are likely associated with Be stars or that have nebular emission. Note that since Br11 line is seen in emission, by convention EqW is negative. \label{fig:EqW}}
\end{center}
\end{figure*} 

To measure the EqW of the Brackett 11-20 lines, we defined windows centered on each line's rest wavelength to measure the line flux from. These windows are defined to be 240 spectral elements (i.e., pixels) in width, such that they typically contain the full width of each Br emission line, regardless of the radial velocity of the star. We then measure the continuum for each line as the median flux of two windows of 121 spectral elements on either side of this line region. Figure \ref{fig:EqW}a highlights these continuum regions with green shading, on either side of the line region for a strong Brackett 11 (which we abbreviate hereafter as Br11) emission line. We measure the pure emission component by subtracting the interpolated continuum flux from the flux in the main line window; integrating the residuals and dividing by the continuum flux provides our EqW measurement in units of \AA.

We calculate uncertainties for these EqW measurements using the formalism of \citet{Vollmann2006}, as validated by a comparison to the observed variance of line measurements for sources with multiple observations. The \citet{Vollmann2006} formalism allows the uncertainty in an EqW measurement to be calculated as:

\begin{equation}
\sigma(W_{\lambda}) = \sqrt{1 + \frac{\bar{F_c}}{\bar{F}} } \frac{ (\Delta \lambda - W_{\lambda}) }{S/N}
\end{equation}
\noindent where $\bar{F_c}$ and $\bar{F}$ are the mean flux in the continuum and line regions, respectively, while $W_{\lambda}$, $\Delta \lambda$, and S/N are the EqW, actual width, and mean S/N of the line in question. Of these values, all but the line's actual width are calculated automatically during the EqW measurement. The line's actual width is more difficult to measure in an automated fashion, so we approximate $\Delta \lambda$ as the full width of our line measurement window. As typical lines only span half to a third of the line measurement window, this will bias our error estimates toward somewhat larger values than would be calculated with a line-by-line width measurement. The Br11--20 lines share a similar width across the full decrement, however, as the width is largely set by the doppler broadening imposed by the velocity structure of the accretion stream. As a result, while the line-by-line uncertainties will likely be overestimated by a factor of $\sim$2, this overestimate will not bias the chi square minimization we perform in Section \ref{sec:model_fits}, but instead will simply adjust the zeropoint of the chi-square minimization process.

\section{Results}\label{sec:results}

\subsection{Catalog}

\subsubsection{Catalog Construction}\label{sec:catalog}

We consider spectra to have significant emission if Br11 EqW $>$ 0.75 \AA (Figure \ref{fig:EqW}b), and a ratio of the measured EqW to the estimated uncertainty of greater than 1. Restricting our attention to spectra obtained with the Young Cluster Plates listed in Table \ref{tab:yc_plates}, we identify 3195 spectra of 879 sources that meet these criteria for a significant Br11 emission line. We report the line measurements, and their errors, for these sources in a machine readable csv file, whose columns are documented in Table \ref{tab:dr17_data}\footnote{We do null Br12-20 line measurements for four young cluster targets whose line measurements produce spuriously high line or uncertainty values: 2M03423420+3151008, 2M04541715+3142224, 2M05011102+3114125, and 2M10450433-5957352}. Expanding our search to all spectra in DR17, we find 9733 visit spectra corresponding to 4255 unique sources satisfying such a criterion. We present EqW measurements for these sources in the machine readable table whose columns are documented in Table \ref{tab:dr17_data} as well, but the processes producing these Brackett emission lines are much less likely to be related to pre-main sequence magnetospheric accretion. Be stars, for example, are known to exhibit APOGEE spectra with prominent, but often double-peaked, Brackett emission lines \citep[e.g., ][]{chojnowski2015}; we also detect a number of spectra with narrow (full width $<$ 200 km/sec) Brackett emission lines in known H2 regions, which we classify as likely nebular sources. As a result, we do not analyze this larger catalog directly, but use it instead as a resource to refine our sample of bona fide YSO emission line sources, by making manual measurements for a subset of the broader catalog and then training a neural network to identify likely Be stars or nebular emission lines based on the morphology of their Brackett emission line profiles.

\subsubsection{Machine Learning}\label{sec:machinelearning}

To further remove contaminants in an automatic way, and provide more accurate estimates of the line profile parameters such as the line width and single/double peak status, we constructed and trained a neural network using a training set constructed from visual examination of a random subset of 500 spectra of Br emitting candidates. In examining it, we have flagged lines that did not have Br emission, or lines that appeared to be double-peaked \citep[such emission most likely originates from Be stars,][]{chojnowski2015}. For the lines that did have Br emission, we manually measured equivalent width using a custom interactive interface through specifying the apparent full width of the line and the continuum. We also estimated the central velocity of the line. Four people have performed independent manual measurements for the same set of 500 spectra, to characterize systematic differences in the placement of the continuum.  \footnote{In the course of the visual analysis to manually measure line strengths for this training set, we identified 63 sources with emission lines blueward of their Br11 line. We show examples of these spectra in \ref{fig:FeII_emitters} the appendix, and include an 'FeII' column in the main machine readable table to highlight these sources, coded with flags of 1, 2 and 3 respectively.}

Afterwards, we constructed a neural network to to automate the classification. The training set consisted of the 80 \AA\ cutout of the APOGEE spectra around all Br11-20 lines, centered on the rest wavelength of each line. To minimize the noise and optimize the training process, these spectral windows have been rebinned to 0.25 \AA\ resolution, such that each spectral window consisted of 320 separate data points. To prevent the model to base all of the predictions for a given spectrum just based on the strongest Br11 line, each line in each spectrum was treated separately, as were all four independent manual measurements for each line. Thus, the set has consisted of 20,000 spectral windows, consisting of $\log \mathrm{Flux}$ to bring the data to a more narrow range of inputs. In total, 80\% of this set was used for training, and 20\% has been withheld as a validation sample.

The neural network model was constructed using Tensorflow \citep{tensorflow}, consisting of 4 convolutional layers, and 4 fully connected layers, with layers connected with the tanh activation function. Two separate tasks were performed. First, classification, providing a probability that a given spectral window had one of three flags: 1) an emission line is present in the spectrum, 2) an emission line is present, but it appears to be double-peaked, and 3) there is no emission line in the spectral window. The probability of all three would add up to 100\% for each line.

Following the classification, a separate network with a similar architecture would perform regression tasks. For the spectral windows with a detected line, it would estimate the log equivalent width, log full width, and log wavelength offset to the line center.

Sparse Categorical Crossentropy Loss was minimized for training the classification model, and Mean Squared Error Loss was used for the regression model. Both models used Adam optimizer with the learning rate of 1e-4. The training has continued until the loss in the validation sample did not improve for 20 epochs.

After applying the trained model on all spectra identified in Section \ref{sec:catalog}, we identify the bona fide emission lines. Leveraging all of the model predictions for all the lines, we identify the maximum Br line that is detectable in the spectrum. Since the line profile should be very similar for all of the detected lines, we average the classification to identify double-peaked sources. We similarly estimate doppler shift for all of the lines, as well as the typical width of the line in units of km s$^{-1}$. Based on the width of the line profile, we also identify sources most likely to be nebular in origin: such sources tend to have narrow lines with an average full width $<250$ km s$^{-1}$, and Br11 width $<13$ \AA. In contrast, emission lines for the accreting young stars tend to have full width of $\sim$300--400 \AA, and double-peaked Be stars tend to have full width of $\sim$800 \AA.

As noted earlier, our automated EqW measurements identify Br11 emission features in 3195 spectra of 879 unique sources on the young cluster plates listed in Table \ref{tab:yc_plates}. Our trained neural network independently confirms the emission lines detected for $\sim$98\% of these spectra (3136 / 3195) and sources (858 / 895). Of these confirmed emitters, however, the network classifies nearly a third as having line profiles consistent with a Be star classification (1256 spectra of 299 sources) or a spectrum with nebular emission features (779 spectra of 282 sources). The remainder, consisting of 1101 spectra of 326 sources, exhibit single-peaked emission profiles with velocity widths consistent with magnetospheric accretion. Together with their angular proximity to known star forming regions and young clusters, we identify these sources as likely accreting pre-main sequence stars, and proceed to analyze their line strengths to infer the temperatures and densities of their accretion flows.

We report these neural network classification results in the machine readable table whose columns are documented in Table \ref{tab:dr17_data}.

\subsubsection{Comparison to Optical Accretion Rate Estimates} 

\begin{figure}
\begin{center}
\includegraphics[scale=0.5]{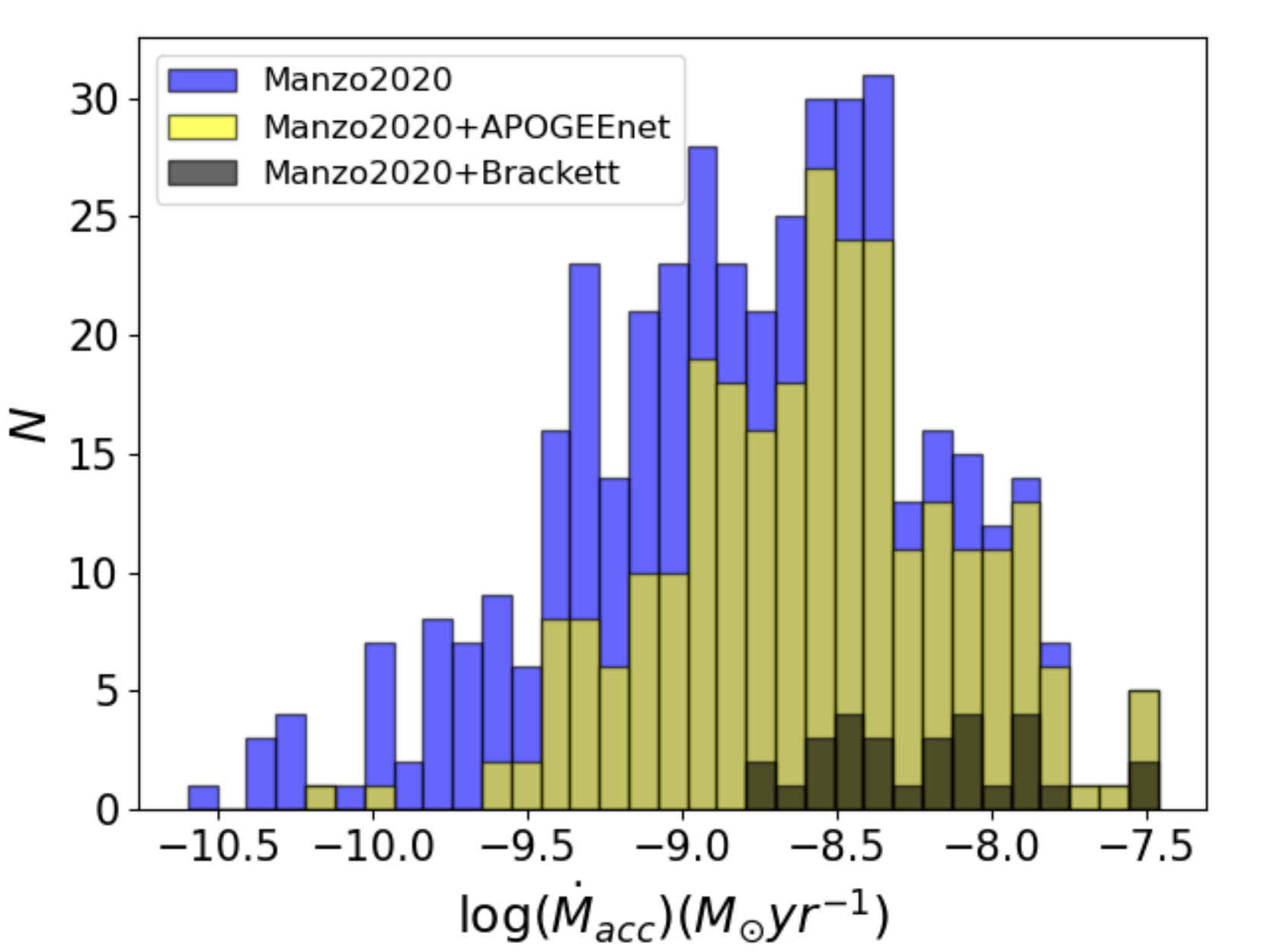}
\caption{Accretion rate measurements for sources with optical measurements by \citet{Manzo-Martinez2020} (full sample shown as the blue histogram), highlighting the subsets of sources with APOGEEnet stellar parameters (yellow histogram), and with Brackett emission lines detected in this work (black histogram). The Brackett sample is strongly biased towards the most rapidly accreting sources in the \citet{Manzo-Martinez2020} sample, given the larger temperatures and densities required to populate hydrogen's n=11-20 levels, and the larger optical depths that are necessary to produce detectable emission lines. \label{fig:Manzo_comparison}}
\end{center}
\end{figure}

To quantify typical mass accretion rates for pre-main sequence stars with robustly detected Brackett emission lines, we analyze accretion rate measurements reported by \citet{Manzo-Martinez2020} based on analysis of optical H$\alpha$ line measurements for 835 T Tauri stars in the ONC, $\sigma$ Ori, Orion OB1a/b, and Taurus. The full sample of 420 optically derived accretion rates measured by \citet{Manzo-Martinez2020} are shown in Figure \ref{fig:Manzo_comparison}, with histograms overlaid to highlight the 268 sources with optically derived mass accretion rates for which APOGEEnet spectral parameters (and thus APOGEE spectra) are available, and the much smaller set of 30 sources for which a Br11 emission line has been detected. As Figure \ref{fig:Manzo_comparison} demonstrates, we only detect Br11 emission for $\sim$11\% of the 268 stars with APOGEE spectra and accretion rates measured by \citet{Manzo-Martinez2020}.  The sources in the \citet{Manzo-Martinez2020} sample with Br11 detections are also strongly biased towards higher mass accretion rates: their optically derived mass accretion rates all exceed the median rate measured by \citet{Manzo-Martinez2020} of 1.6$\times10^{-9}$ M$_{\odot}$/yr.  

Figure \ref{fig:Manzo_comparison} also demonstrates, however, that while all pre-main sequence stars with well-detected Brackett emission lines have high optically derived accretion rates, the inverse does not hold: we do not detect Brackett emission lines from all sources with APOGEE spectra and large optically inferred accretion rates. Of the 268 sources with both APOGEE spectra available and accretion rate estimates by \citet{Manzo-Martinez2020}, 179 (or more than 2/3rds) have an optically derived accretion rate larger than the smallest accretion rate measured for the 30 sources with Br11 detections.  Put another way, we only detect Br11 emission for 1/6th of the sources with accretion rates higher than the lowest Br11 detection: 5/6ths of these high accretion rate sources are null detections for our Brackett analysis. To better understand the nature of the sources with strong H$\alpha$ derived accretion rates but no detected Brackett emission lines, we visually inspected all APOGEE spectra available for the 30 sources with the highest optical accretion rate measurements in the \citet{Manzo-Martinez2020} sample, but which were not identified as strong accretors in the Brackett emission sample. APOGEE spectra are available for 24 of these 30 high accretion rate sources, but none have Brackett line detections reported in the machine readable table whose columns are documented in Table \ref{tab:dr17_data}.  Visual inspection confirms a lack of reliable Brackett line emission in the APOGEE spectra of these 24 sources, particularly in the Br11 line used to select sources with robust emission detections.  Three sources (2M05343822-0524236, 2M05345292-0528591, 2M05352463-0519096) appear to exhibit Br11 emission just below the detection limit of our catalog, but their Br11 EqWs and/or SNR estimates do lie just outside the bounds of our detection limit, such that the Br11 emission feature must be considered as marginal rather than secure. We comment more on the astrophysical implications of the disconnect between these optical and near-infrared accretion indicators in section XXXX, limiting the discussion here to verifying and documenting the lack of agreement between Balmer and Brackett selected accretors.

\subsection{Model Fits} \label{sec:models}

\subsubsection{Kwan and Fischer (2011) Line Strength Calculations}

\citet{KwanFischer2011} performed detailed local excitation calculations to determine the emissivity ratios of Hydrogen lines in the Balmer, Paschen, and Brackett series. Unlike prior work which makes strong assumptions about the optical thickness and level populations of the line ratios in question \citep[e.g., the Case B recombination models calculated by ][]{Baker1938, Hummer1987}, \citet{KwanFischer2011} calculate the radiative and collisional equilbria for each set of input parameters, and calculate level populations and emissivities in a self consistent manner. These calculations were performed assuming a nominal UV excitation field, over a range of temperatures (5000 $< T <$ 30,000 K) and hydrogen nucleon number densities (10$^8 < n_H <$ 2 $\times 10^{12}$ cm$^{-3}$) relevant for gas in protostellar accretion flows. In their initial work, \citet{KwanFischer2011} reported emissivity ratios for Hydrogen lines up to n=15; as APOGEE spectra sample transitions up to the Br20 line, additional calculations of the emissivity ratios of these higher order transitions were kindly provided by Kwan \& Fischer (2015, private communication) for the same grid of temperatures and densities as in their 2011 publication. \footnote{We note that the calculations for line ratios up to the n=20 state of hydrogen are now publicly available at \url{https://www.stsci.edu/~wfischer/line_models.html}}

\begin{figure}
\begin{center}
\includegraphics[width=\columnwidth]{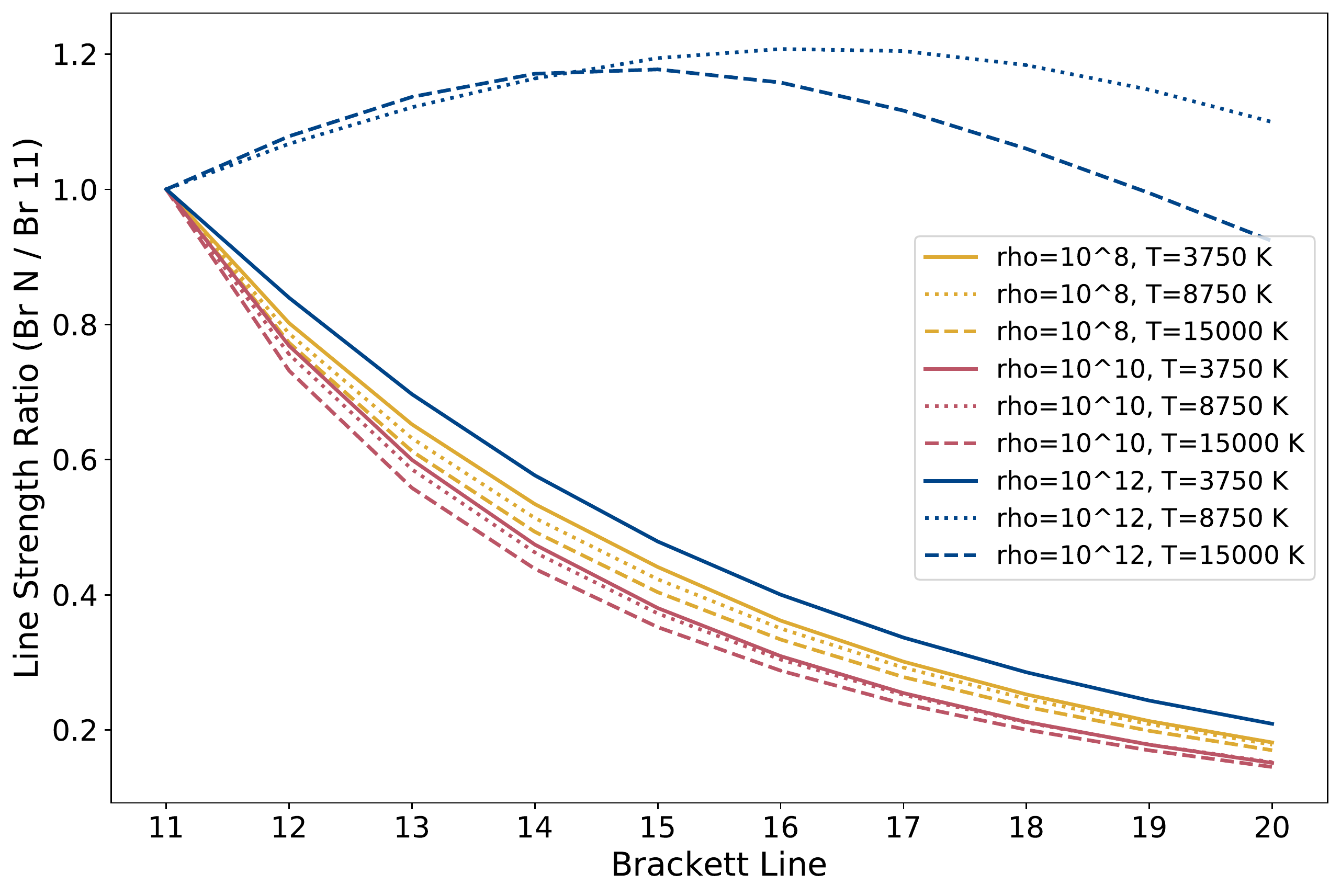}
\caption{\citet{KwanFischer2011} model grid, showing the decrement of the emission in Brackett lines relative to the Br11 line, separated into a range of different temperatures and densities of the emitting gas. Note that the most distinct decrement models have high density and high temperature. \label{fig:model_grid}}
\end{center}
\end{figure} 

The relative line strengths predicted for a sparse, grid-spanning set of temperatures and densities are displayed in Figure \ref{fig:model_grid}. For each model, all line strengths are normalized by the observed strength of the Br11 line; we refer to a set of such normalized line strengths, either in our model grid or as measured from an empirical spectrum, as a Brackett decrement. At lower densities (i.e, $n_H \sim 10^8$ and $10^{10}$), the Brackett lines are optically thin, and the decrement is only marginally sensitive to the gas temperature. As such, even moderate observational errors may significantly affect the confidence of the temperature determinations for any targets whose decrement is consistent with a lower density fit. The decrements predicted for higher density models are significantly more temperature sensitive, however, and populate a much more unique area of line ratio parameter space. As a result, in the absence of substantial observational errors, any fits that return a higher density will also provide a much more precise constraint on the temperature of the emission region.

\subsubsection{Calculating and Fitting Empirical Brackett Decrements}\label{sec:model_fits}

Once we calculated the EqWs of the Brackett 11-20 lines for each spectrum, we took the ratio of each spectrum's line EqWs to their respective Br11 EqW to measure the decrement in line strengths from Br11 to Br20. We then compared these measured decrements to the predictions by \citet{KwanFischer2011} for the 161 temperature and density points in their model grid.  This grid spans densities from $10^8$ to $10^{12.4}$ cm$^{-3}$ and temperatures from 3750 K to 15000 K. 
 
It is important to note that the decrements we construct from continuum normalized EqW measurements are not fully equivalent to the flux-weighted intensity ratios predicted by the \citet{KwanFischer2011} model grid.  Producing astrophysically accurate intensity ratios would require not only flux calibrated spectra, which APOGEE provides, but also accurate corrections for the effects of foreground extinction on both the line and continuum flux.  As the line flux and continuum sample different emission regions (i.e., the accretion column and stellar photosphere, respectively), determining a robust extinction likely requires detailed modelling of the circumstellar environment.  We have estimated the uncertainties associated with fitting Br decrements based on EqW measurements, rather than extinction corrected line fluxes, however, by examining the range of spectral slopes in our sample and the additional wavelength dependence imposed by uncorrected foreground extinction.  The median spectral slope in our YSO sample, measured as the ratio of the continuum fluxes for the Br11 to Br20, is 0.9, with 10\% and 90\% deciles of 0.78 and 1.16, respectively.  That is, our median EqW-based Br20/Br11 line strength ratio is underestimated relative to the intensity-weighted ratio by 10\% (i.e., a measured EqW ratio of 0.2 would correspond to an intensity-based ratio of 0.22).  This bias scales with the wavelength separation of the lines in question, so will be lessor for line ratios that have smaller separations (ie, the bias in the Br13/Br11 ratio should only be half as large), and is typical a factor 2-3 lower than the uncertainties in our line ratios.

The uncertainty in our EqW based line ratios due to neglecting effects of extinction is harder to scope. Extinction of A$_V$= 1.5 will skew the continuum slope across the APOGEE spectrum of a T $\sim$4000 K source by about 4\%. If the line formation region and the stellar photosphere have the same foreground extinction, however, neglecting this correction should not affect the accuracy of our EqW based line ratios, as both the line and continuum regions will be equally extinguished, and the the EqW ratio will correspond directly to the intensity weighted ratio.  However, it is not obvious that the Br line formation region, which likely samples some of the accretion column as well as the accretion shock, will have the same line-of-sight extinction as the stellar-disk-averaged photospheric flux, which will dominate the continuum.  Indeed, one might expect the photosphere to have a somewhat higher foreground extinction than the line formation region, such that the EqW-based ratios will be biased towards higher values than ratios based on intensities, as the continuum will be preferentially surpressed for the bluer lines in the decrement.  Accounting for these extinction effects would likely require detailed modelling of the circumstellar environment of each source, as the geometry and temperature structure of the circumstellar material will play a role in determining the amount of dust that is present along the line of sight to the different components that contribute to the spectrum.

Given the scale of these effects relative to our existing line measurement uncertainties, and the potential that the biases due to neglecting the measured spectral slopes and potential line-to-continuum extinction differences at least partially offset one another, we have not attempted to convert our EqW-based line ratios into astrophysically corrected intensity ratios.  However, to allow the reader to assess the potential impact of performing these fits in EqW rather than intensity space, we include our Br20-to-Br11 spectral slope measurements in the machine readable table whose columns are documented in Table \ref{tab:dr17_data}.

We used a least squares regression to find the best density and temperature model for each spectrum. We calculated the total RMS difference between the line strengths observed in each decrement and and each model in our decrement grid, after normalizing each difference by the error associated with the measured line strengths, and then selected the model profile with the smallest total normalized difference as our best-fit model for that decrement. Two constants were added to the chi-squared calculations to create a minimum error in the normalized EqWs: a base, flat error of 0.01 to provide a minimum error level that accounts for effects whose precision does not scale cleanly with SNR, such as OH line subtraction and telluric absorption correction; and an error that scales with 2\% of the ratio between the emission line and the Br11 EqW, to account for potential systematic errors that do not scale with SNR, such as the measurement of the stellar continuum. 

These fit results are presented in the machine readable table whose columns are documented in Table \ref{tab:dr17_data}.

\begin{figure*}
\begin{center}
\plottwo{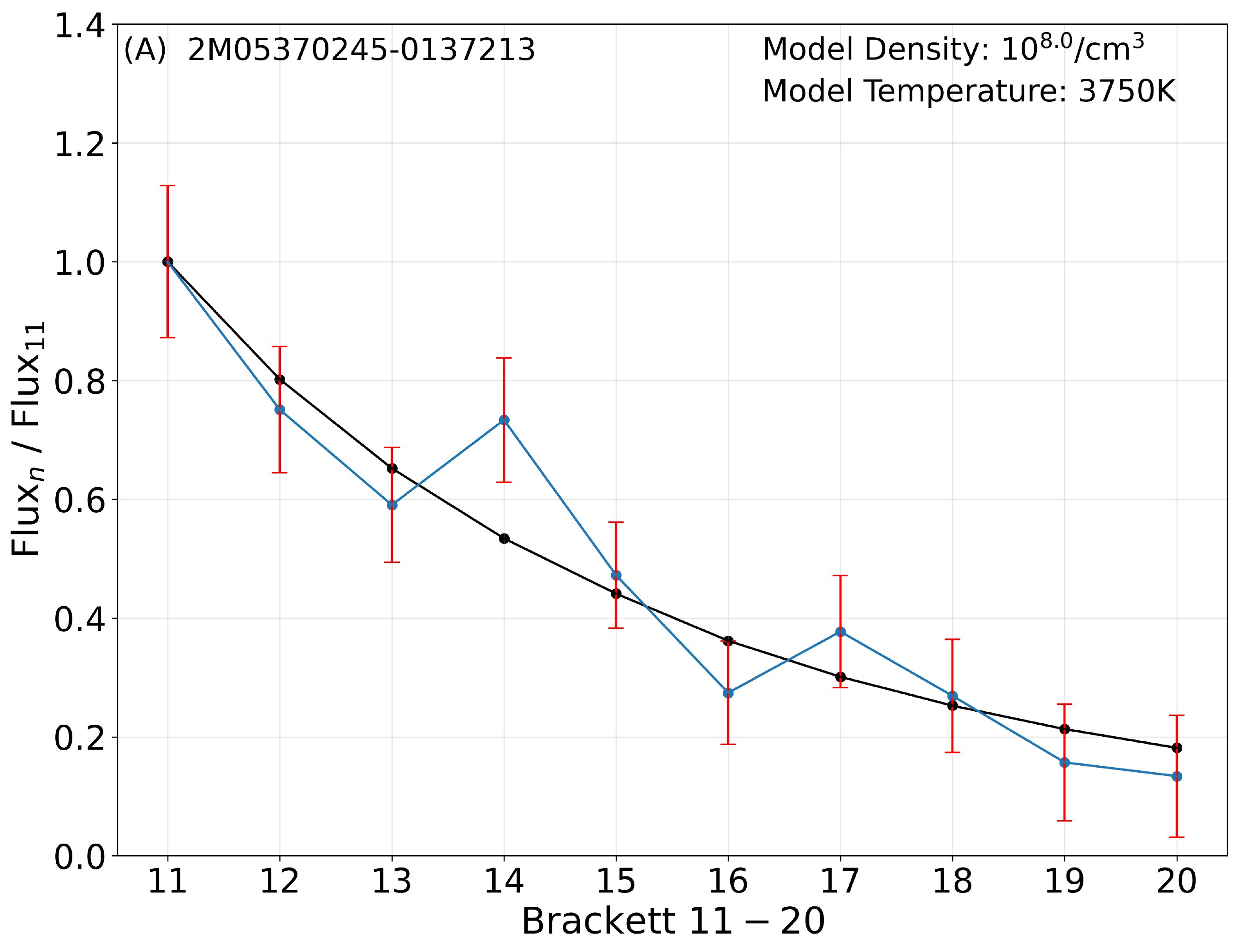}{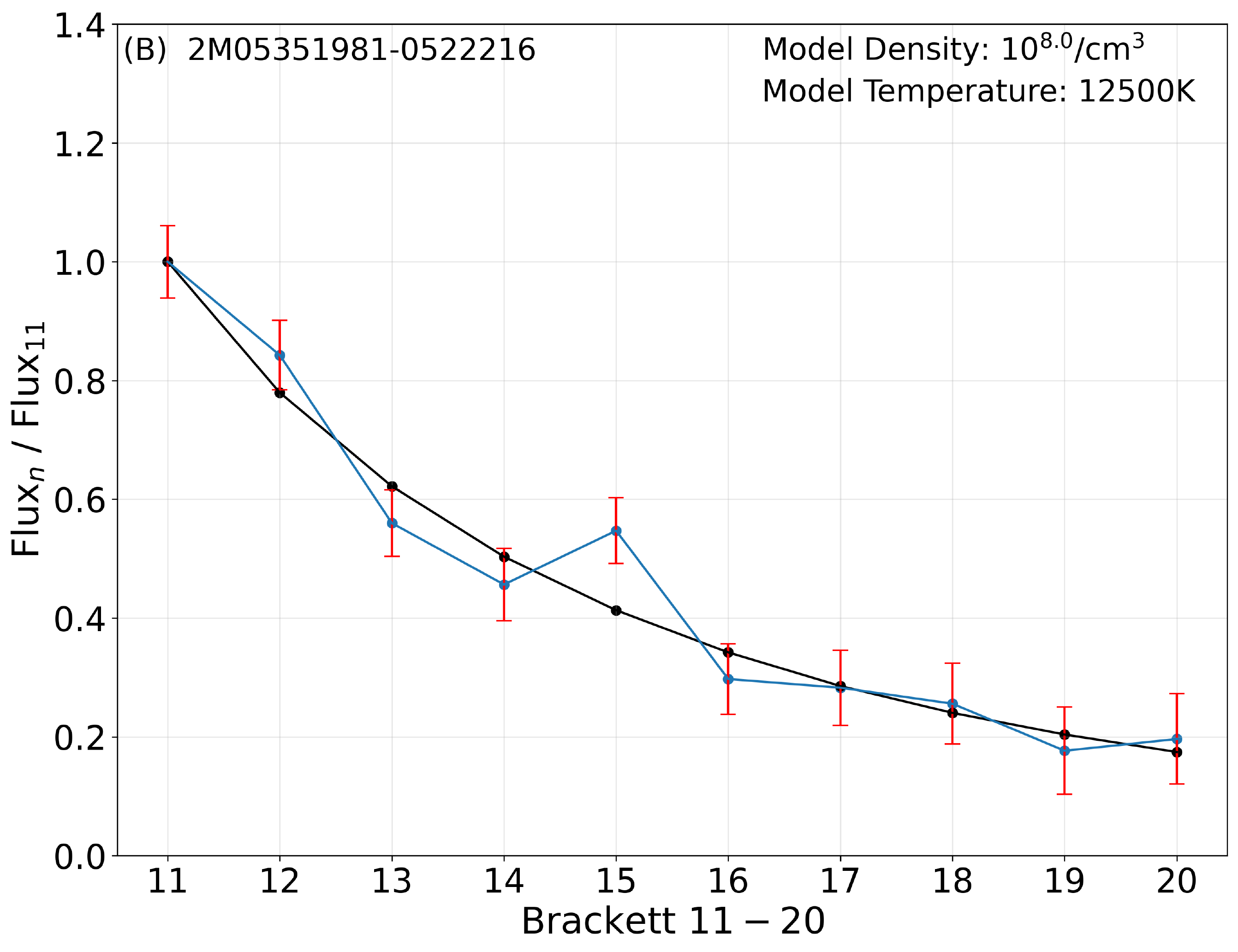}
\plottwo{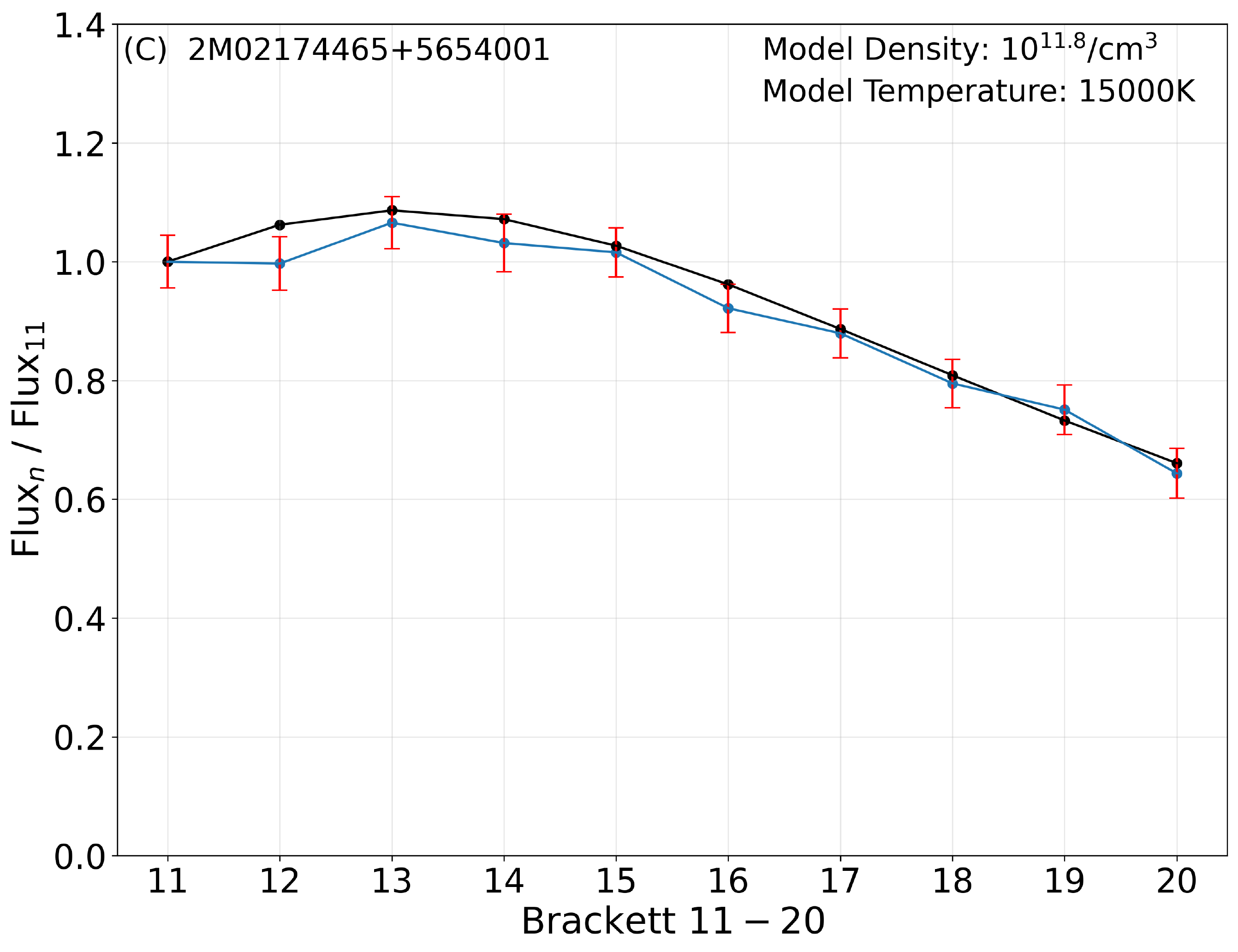}{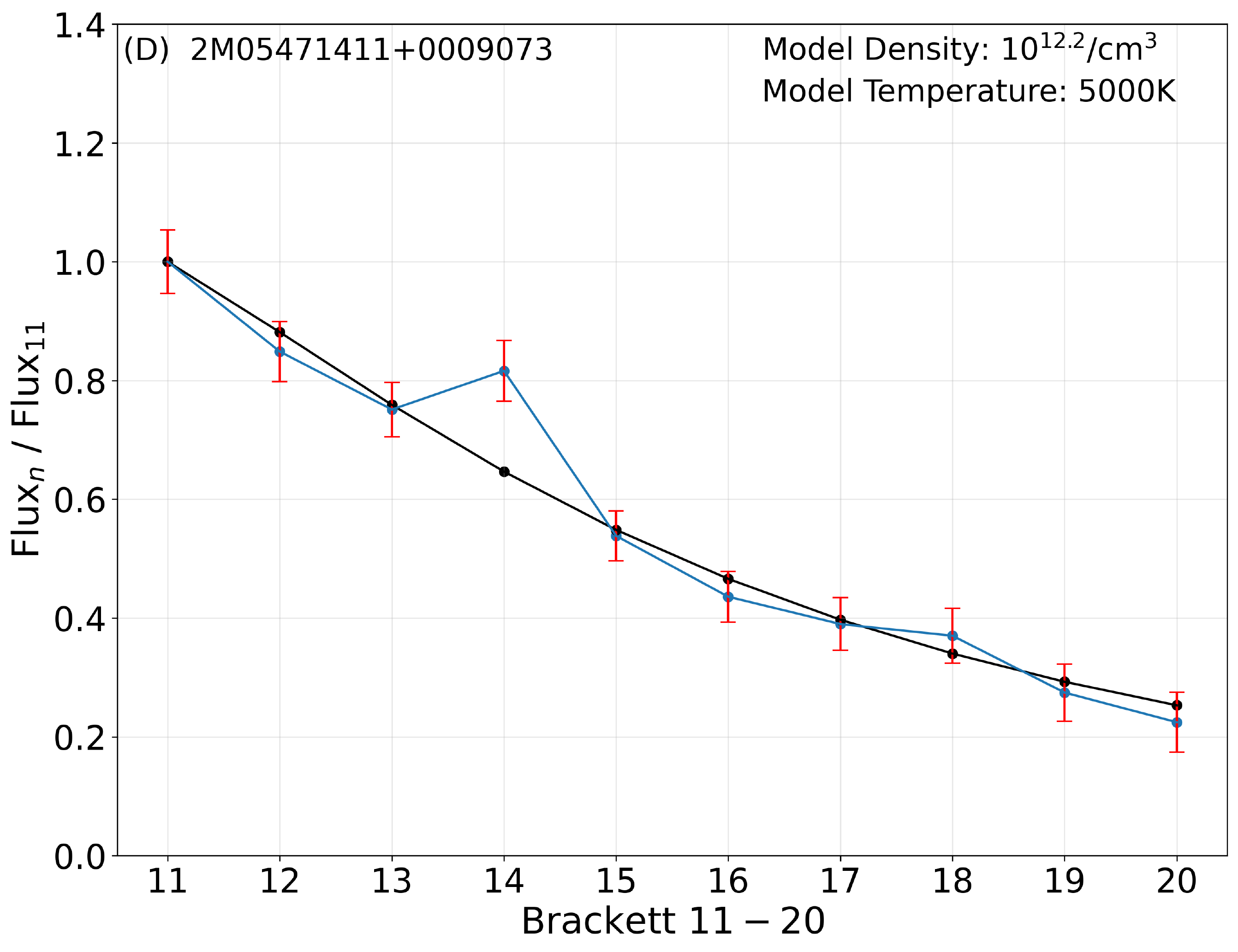}

\caption{Sample decrement fits for several APOGEE spectra from the Young Cluster plates, ordered by density. Each spectrum shows a unique combination of model density and temperature values. The empirical decrement measured from the spectrum is shown as red data points, with empirical errors, while the best fit model decrement is shown as a solid black line. Notes on the individual sources: a) Br 14 is located close to the chip gap, resulting in an uncertain estimate of the continuum. b) Very narrow lines, likely nebular in origin, Br15 is affected by the telluric subtraction. c) Asymmetric double peaked profile, likely Be star. d) Br14 appears to be affected by the blending with Si I 15892.771 line that is only rarely seen in emission.}
\label{fig:sample_decrement}
\end{center}
\end{figure*} 

\begin{figure}
\begin{center}
\includegraphics[scale=0.4]{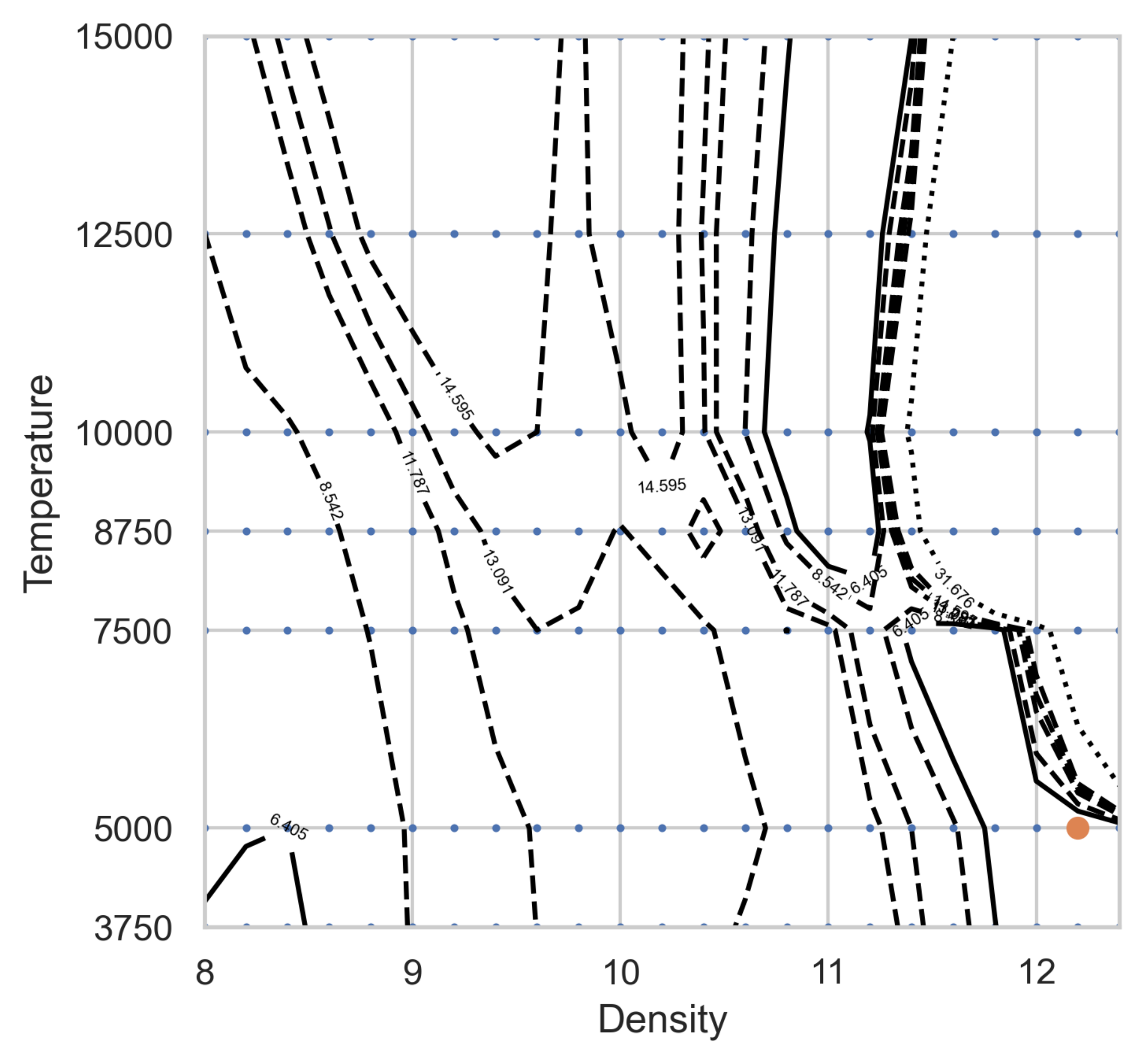}
\caption{Visualization of the chi square surface for the decrement measured from the spectrum of 2M05471411+0009073 taken on MJD = 57433 (8890-57433-87). The orange dot shows the best fitted model.\label{fig:sample_chi}}

\end{center}
\end{figure} 

\subsubsection{Distribution of Best Fit Temperatures and Densities} 

\begin{figure}
\begin{center}
\includegraphics[scale=0.6]{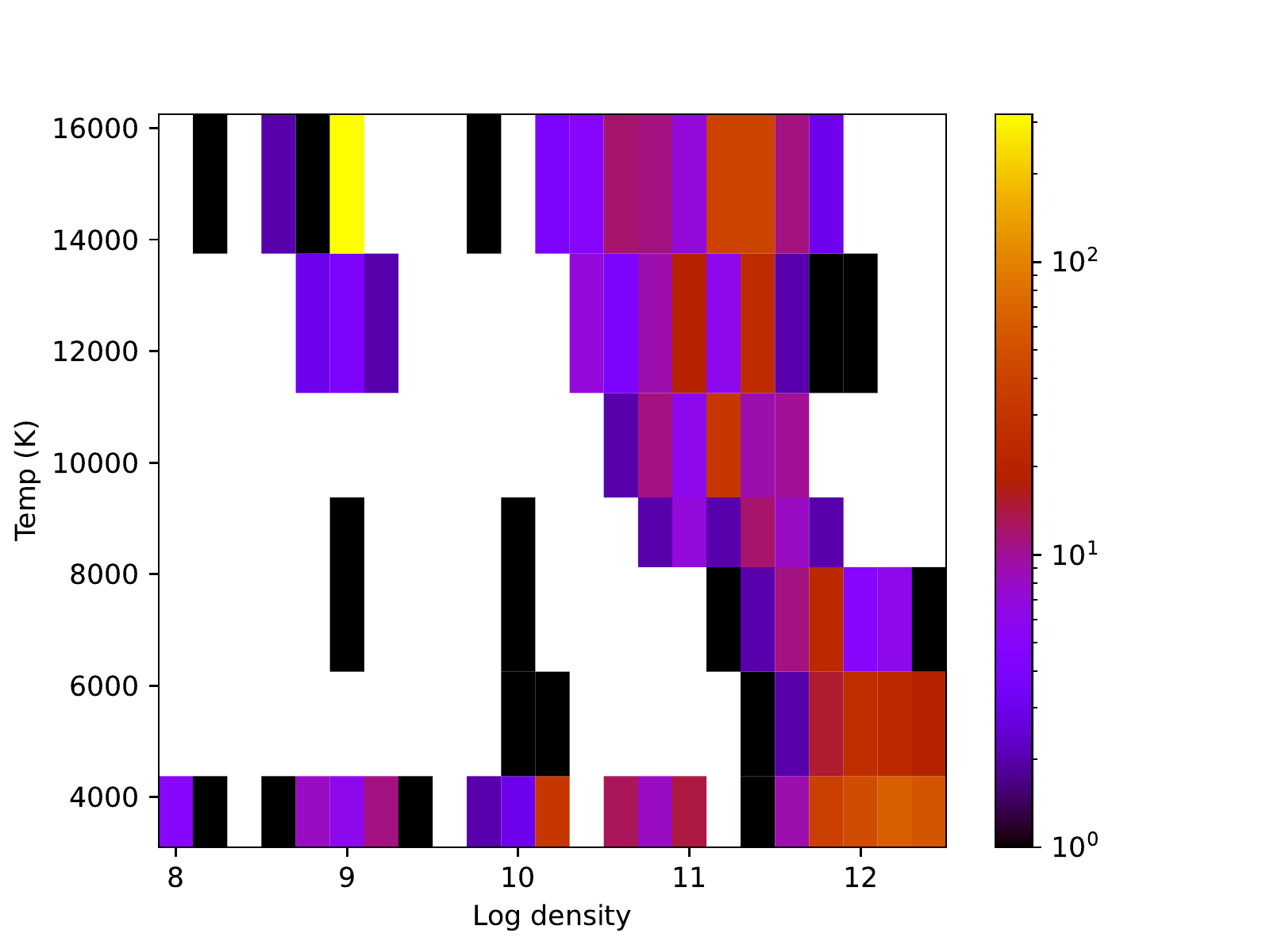}
\caption{Distribution of model fits to 1101 spectra of 326 pre-main sequence accretors in temperature / density space. \label{fig:model_fits}}

\end{center}
\end{figure}

Figure \ref{fig:model_fits} shows the distribution of temperatures and densities identified from fitting the theoretical Brackett decrements calculated by Kwan \& Fischer to the empirical decrements measured from the 1101 spectra of 326 pre-main sequence accretors identified in Sec. \ref{sec:machinelearning}. Consistent with the chi square surfaces of individual sources, such as that shown in Figure \ref{fig:sample_chi}, the fit results indicate a clear preference for a loci of models stretching from $n_H \sim 10^{11}$ and T=15000 K to $n_H \sim 10^{12.2}$ and T = 3750 K. This loci tracks a region of degeneracy in the Kwan \& Fischer model grids, where increases in density above n$_H \sim 10^9$ result in mid-decrement lines, such as Br15, strengthening relative to Br11, while increases in temperature cause the same mid-decrement lines to weaken. As a result, changes in the morphology of the decrement are, to first order, degenerate between anti-correlated changes in temperature and density: relative to a model with $n_H$ 10$^{11}$ and T=10000 K, a similar fit quality will be achieved by a model with somewhat higher density and somewhat lower temperature, allowing the competing influences on the resultant decrement morphology to roughly cancel out. Possibly due in part to the presence of this degeneracy, the best fit results of the decrement fits span the full range of temperatures sampled in the Kwan \& Fischer model grid, providing no clear indication of the typical temperature structure of pre-main sequence accretion flows, or at least those that produce prominent higher order Brackett emission lines.

The fit results do, however, clearly rule out most low-density flows; indeed, the only density lower than $n_H \sim$10$^{10}$ that is favored by a number of best fit models is the grid point associated with $n_H \sim$10$^{9}$ and T$=15000$ K. This set of model parameters represents the edge of the grid in decrement space, where the strengths of mid-level Brackett lines (e.g, Br15) decline most significantly relative to the strength of Br11. As a result, any empirical decrement where the line strengths drop off more steeply than the Kwan \& Fischer models suggest will find their best fit with this theoretical decrement, even if the fit is not particularly good in a global sense. We therefore interpret the 299 sources whose best fit properties correspond to this grid point as likely spurious in nature, indicating a set of conditions, or potentially viewing angles etc, which do not correspond well to a single-parameter solution as required by this current grid comparison. 

\section{Discussion} \label{sec:disc}

\subsection{Accretion Physics}

Observational support for the magnetospheric accretion model has increased significantly over the past three decades, but the initial theoretical calculations underpinning the paradigm remain quite relevant today.  \citet{Hartmann1994} provided the first detailed models of the physical properties of magnetospheric accretion streams, predicting electron densities of $n_e \sim 10^{10} - 10^{12}$ and temperatures of 5500-7000 K along streamlines connecting the surface of a typical 0.5 M$_{\odot}$ T Tauri star to its circumstellar disk.  These models predict that the electron densities, and the densities of bound hydrogen atoms in the second excited state, should steadily increase along each streamline as the accreting material approaches the stellar surface.  Due to the enhanced efficiency of radiative cooling mechanisms at higher densities, however, the temperature predicted along each streamline does not increase monotonically, but instead rises quickly from 5500K to 7500K shortly after the streamline departs the stellar disk, and then slowly decreases back to 7000K as the streamline approaches the stellar surface. As a result, these models suggest material in magnetospheric accretion columns should exhibit characteristic densities of $n_e \sim 10^{10} - 10^{12}$, with higher opacity line tracers preferentially sampling lower densities further out in the streamline, and lower opacity tracers sampling higher densities and a narrower range of excitation temperatures close to 7500K.  

Subsequent works have explored the complexity of the geometry and radiative processes in accretion flows, but have not fundamentally revised the temperature and density structures assumed in theoretical models. \citet{Calvet1998} described the structure of the accretion shock and post-shock regions implied by these models, while \citet{Muzerolle2001} provided improved calculations of the emergent line emission, and identified that the maximum temperature consistent with observed line profiles appears to be inversely correlated with accretion rate (i.e., $T_{max} > 10^{4} K$ is allowed for accretion rates of $\dot{M} < 10^{-9} M_{\odot}/yr$, but $T_{max} > 7000 K$ is prohibited for $\dot{M} \gtrsim 10^{-7} M_{\odot}/yr$; see their Fig. 16).   Computational advances have also allowed significantly more complex radiative transfer methods to compute emergent line profiles \citep[see, e.g., ][]{Kurosawa2011, Esau2014, Dmitriev2019, Wilson2022}, and point to more complex geometries than implied in a simple dipole model \citep[e.g., ][]{Romanova2003, Romanova2008, Ingleby2013}. Nonetheless, the formalism used by \citet{Hartmann1994} and \citet{Muzerolle2001} appears sufficient to describe the density and temperature structure of the accreting material in even non-dipolar flows and provides significant explanatory power for modern observations. The 3-D numerical model computed by \citet{Esau2014} for AA Tau, for example, predicts that the density of the accretion flow increases by two orders of magnitude from the circumstellar disk to the stellar surface, while producing a hot spot on the stellar surface of 5500 K. 

Given the stability in the temperatures and densities expected to dominate the emission line regions of magnetospheric accretion columns, the densities we infer from our Brackett decrement observations can be compared to those measured by a suite of complementary analyses based upon the \citet{KwanFischer2011} grid.  \citet{Edwards2013} modelled Paschen decrements observed for a representative sample of CTTSs, finding best-fit densities of $n_H \sim 10^{11}$.  \citet{Rigliaco2015} infer similar densities of $n_H \lesssim 10^{11}$ from ratios of mid-infrared Pfund HI lines measured from Spitzer spectra of CTTS and transition disk sources.  With the broad optical and infrared coverage offerred by X-Shooter spectra, \citet{Antoniucci2017} model Balmer and Paschen decrements for CTTSs spanning a range of accretion rates, finding best fit densities of $n_H > 10^{11}$ for their highest accretion rate sources, but lower densities ($n_H > 10^{9.6}$) for sources with lower accretion rates and/or for fits to the higher opacity Balmer decrements. The range of tracers and accretion rates sampled by \citet{Antoniucci2017} demonstrate that higher optical depth tracers sample gas from a broader region of the magnetosphere, including lower density gas that resides higher up in the accretion column. 

The higher densities ($n_e > 10^{11}$) favored by our decrement fits, as well as the relative paucity of Brackett line detections among the optically selected strong accretors from the \citet[][see Figure \ref{fig:Histograms}]{Manzo-Martinez2020} sample, can both be understood as the result of optical depth effects predicted by the \citet{KwanFischer2011} model grid. The \citet{KwanFischer2011} calculations indicate that the upper level Brackett lines become optically thick at densities of log $n_H\sim$11.  (see, e.g., the optical thickness of Paschen beta in Figure 11 of \citep{KwanFischer2011}, or the Case II model data files on Will Fischer’s website: \url{https://www.stsci.edu/~wfischer/line_models.html}). At densities greater than 10$^{11}$, the models predict a rapid increase in the relative strength of the higher order Brackett lines, particularly in cases where the excitation temperatures are 8750K or larger. In this regime, the Br11 line is predicted to become comparable in strength to the Br Gamma line, and for the highest densities and temperatures, Br11 will become notably stronger (ie, Br11 will be twice as strong as BrGamma for T=10,000K and log n\_H ~ 12).
 
A similar dynamic explains the relatively poor overlap between our sample and the H$_{\alpha}$ selected accretors identified by \citet{Manzo-Martinez2020}. At low densities, H$_{\alpha}$ is predicted to be 10$^2$-10$^4$ stronger than both Br Gamma and Br 11 (e.g., for log n\_H = 9.0 and T = 8750K, H$_{\alpha}$ is predicted to be 2000x stronger than Br Gamma, and 13,000x stronger than Br11).  As the density increases, however, the ratio drops significantly: at log n\_H = 11 and T = 8750 K, H$_{\alpha}$ is predicted to only be 47 and 60 times stronger than BrG and Br11. At log n\_H ~ 12 and T = 8750 K, ratios have dropped to 12.5 and 6.5, respectively (and the Br11 line has indeed become stronger than Brackett Gamma).  These energetics arguments suggest that physical properties of the accretion streams, and their densities in particular, may be the primary factor determining whether the high accretion rate sources detected by Manzo-Martinez et al. have detectable Brackett lines or not.  That is, the 5/6ths of the H$_{\alpha}$ selected sources that have no detectable Brackett lines may have characteristic accretion stream densities below 10$^{11}$, such that their optical emission lines are easily detected but their higher order Brackett lines remain undetectable.  The remaining 1/6th of sources, which have strong Brackett lines as well as strong H$_{\alpha}$ emission, would then be those sources whose accretion streams have characteristic densities in excess of 10$^{11}$.  In this interpretation, the 1/6 to 5/6 ratio of Brackett detections in the optically selected sample would reflect the relative frequency of accretion streams with densities above and below 10$^{11}$.  Likewise, the consistent finding of densities at or above 10$^{11}$ from most sources with secure Brackett detections would similarly support the picture in which Brackett lines will only become detectable once densities exceed the 10$^{11}$ threshold, and the lines become optically thick.
 
Finally, similar threshold effects could well explain the range of characteristic densities identified in accretion studies using different tracers.  In each case, the mean characteristic density would likely be near, but somewhat above, the density threshold that corresponds to the transition to the line being optically thick.

\subsection{Intrinsic Variability}\label{sec:variability}
\begin{figure*}
\begin{center}
\includegraphics[scale=0.6]{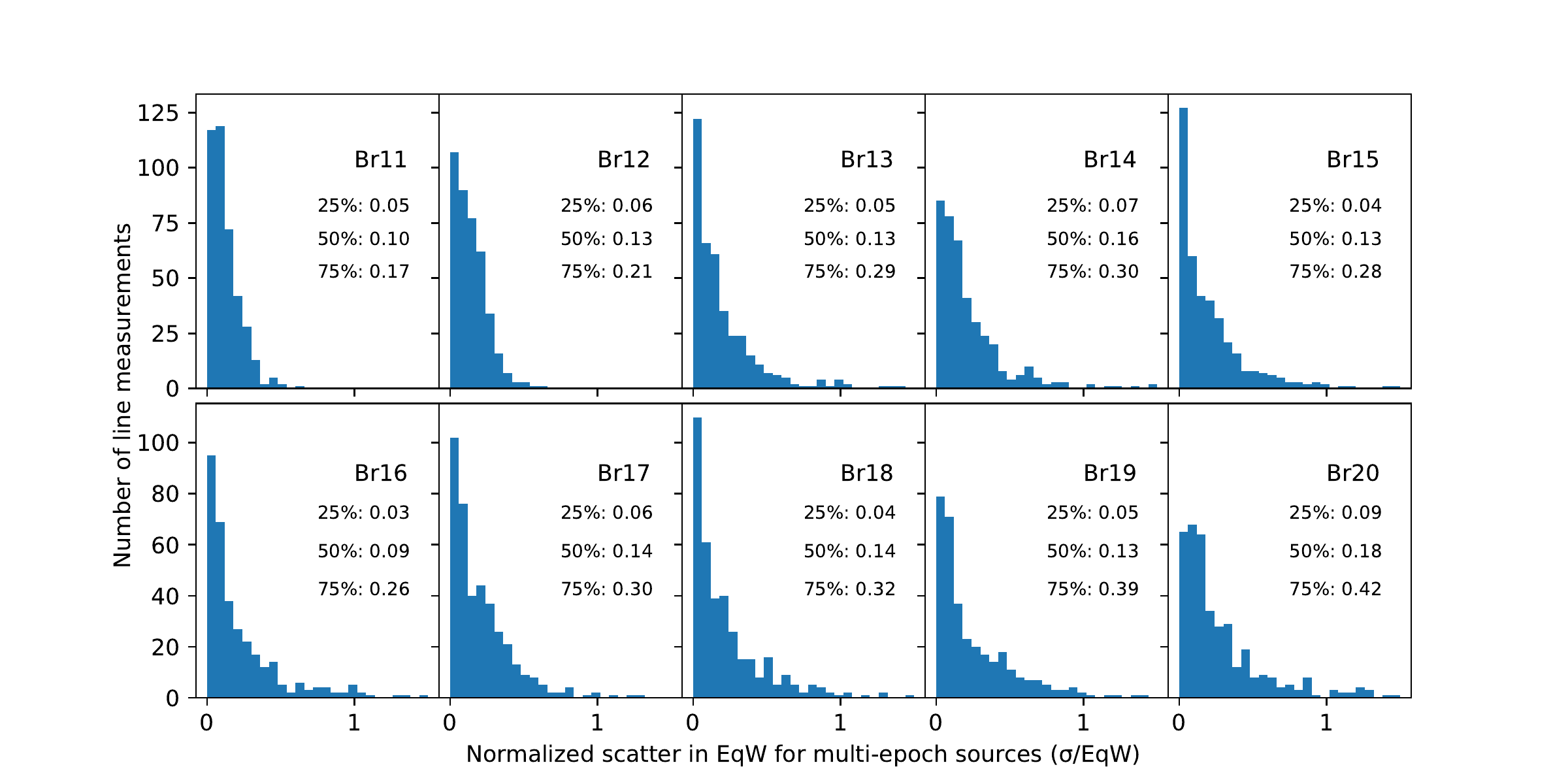}
\caption{Histogram showing the scatter of the Brackett line strengths measured for sources with multiple spectra taken over time, relative to their mean measured line strengths. 25th, 50th, and 75th percentiles of the distribution are included in each panel, and indicate most Brackett emission lines vary by only 10-20\% on timescales of months to years. Some sources exhibit variations as large as 100\%, though more frequently in the higher order lines, whose lower absolute strengths lead to larger fractional variations at fixed signal-to-noise.\label{fig:Histograms}}
\end{center}
\end{figure*}

Some YSOs were observed multiple times on different days, where each observation is called an epoch. We analyse the variability in the measurements of YSOs at different epochs to ensure confidence in the EqW measurements. If the EqW measurement of one YSO varies considerably, the model may fit the YSO to widely different density and temperatures for different epochs. Figure \ref{fig:Histograms} shows how much the different epochs of the Br 11 – 20 lines for each YSO varies. Each data point represents one YSO. The bins are the standard deviation of the different EqW measurements of one YSO normalized by the average EqW of that Br line of that YSO. We see that the majority of the Br lines of the objects with multiple epochs vary by 10 – 20 percent of the average EqW of the Br line for that object. This means the EqW measurements of different epochs of a YSO varies relatively little and therefore we can have confidence in the model fits of the different epochs of the same object.

\section{Conclusions} \label{sec:conclusions}

In this paper we present an analysis of the Brackett emission lines in the APOGEE spectra of young stars.

\begin{enumerate}
    \item Using an automated pipeline that integrates the flux over a pre-defined spectral window, we identify Br11 emission lines in 9733 spectra of 4255 sources in the APOGEE DR17 dataset. For each of these sources, we measure and report in a machine readable table, whose columns are documented in Table \ref{tab:dr17_data},  EqWs and errors for the full set of Br11 – Br20 emission lines that fall within the reach of the APOGEE detectors. 
    \item While there is some contamination in the above selection, we perform a manual classification and EqW measurement of a subset of these Br emission line sources, and train a neural network to more robustly verify the presence of the line, as well as to classify double-peaked line profiles, which are likely to be Be stars. We also identify sources with narrow lines that are more likely to be consistent with the nebular emission.
    \item Restricting our sample to sources with strong, single-peaked Br11 lines, and even further to plates specifically targeting known star forming regions and young clusters, we identify 1101 spectra of 326 likely pre-main sequence accretors.
    \item For each of the 1101 spectra of likely pre-main sequence accretors, we calculate empirical Brackett decrements and fit them with the the radiative transfer models computed by \citet{KwanFischer2011}. For nearly a third of these spectra (320 out of 1101), we measure steeper decrements (i.e, with stronger Br11 and weaker higher order lines) than predicted by any model in the current grid. For these spectra,  the closest matching predicted decrement in the model grid typically corresponds to the density of $n_H=10^9$ cm$^{-3}$ and temperature of T$=15000 K$. The remaining  781 spectra with shallower decrements (that do fall within the bounds of the model grid) show a clear preference for typical densities of 10$^{11}$-10$^{12}$  cm$^{-3}$. Temperatures are less well constrained, but there is a slight preference for fits with temperatures in the range of 8000-15000 K, particularly for spectra with densities near 10$^{11}$  cm$^{-3}$.
    \item The densities indicated by these best fits appear consistent with predictions of accretion stream models \citep{Hartmann1994, Muzerolle2001, Dmitriev2019} and other observational studies that compare hydrogen line strengths to predictions of the \citet{KwanFischer2011} grid to infer densities of accretion columns \citep[e.g., ][]{Edwards2013, Rigliaco2015, Antoniucci2017}. Taken together, these studies support a picture where the density increases along typical CTTS accretion streams from $n_H \sim 10^{9.6}$ \citep[as indicated by the highest opacity Balmer decrements][]{Antoniucci2017} to $n_H \sim 10^{11}$ \citep[as traced by the higher opacity Paschen and Pfund lines][]{Edwards2013, Rigliaco2015, Antoniucci2017}, and reach values as high as $n_H \sim 10^{12}$ for the densest regions of the most heavily accreting systems, as revealed by the lowest-opacity Brackett decrements reported here. 
    \item Brackett line strengths measured for sources with more than one APOGEE spectrum indicate that typical epoch-to-epoch variations are no larger than 10 – 20\%, at or below the level of our typical measurement errors. 
    \item The multiplex capacity of the APOGEE spectrograph and the operational efficiency of the SDSS observing systems were critical for obtaining the observational dataset that enabled the construction of this catalog of Hydrogen emission line decrements, one of the largest assembled for pre-main sequence stars and protostars. SDSS-V has already begun observations of an even larger sample of 100,000 young stars, many of which are expected to show strong accretion signatures. With increased flexibility to co-observe with the APOGEE and BOSS spectrographs, SDSS-IV will obtain both optical and NIR spectra of these young stars, enabling at least partial coverage of the Brackett, Balmer, and Paschen series, as well as the Calcium H\&K and infrared triplets. Building on the analysis presented here, measurements of the line strengths and profiles captured in this rich observational dataset will enable improved constraints on the properties of accretion flows in pre-main sequence stars.  \\\\
    
\end{enumerate}

We thank John Kwan and Will Fischer for their work to calculate and distribute the line ratio predictions that make this analysis possible.  We also thank the anonymous referee for a prompt and useful report which improved the analysis presented here.

H.C. \& E.K. acknowledge support provided by Chandra Award Number GO9-20006X issued by the Chandra X-ray Center, which is operated by the Smithsonian Astrophysical Observatory for and on behalf of the National Aeronautics Space Administration under contract NAS8-03060.

M.K. \& K.R.C. acknowledge support provided by the NSF through grant AST-1449476, and from the Research Corporation via a Time Domain Astrophysics Scialog award (\#24217). 

C.R.Z. acknowledges support from projects UNAM-DGAPA-PAPIIT 112620 and CONACYT CB2018 A1-S-9754, Mexico.

K.P.R. acknowledges support from ANID FONDECYT Iniciación 11201161. 

A.S. gratefully acknowledges support by the Fondecyt Regular (project
code 1220610), and ANID BASAL projects ACE210002 and FB210003.


Funding for the Sloan Digital Sky Survey IV has been provided by the Alfred P. Sloan Foundation, the U.S. Department of Energy Office of Science, and the Participating Institutions. SDSS-IV acknowledges
support and resources from the Center for High-Performance Computing at
the University of Utah. The SDSS web site is www.sdss.org.
SDSS-IV is managed by the Astrophysical Research Consortium for the 
Participating Institutions of the SDSS Collaboration including the 
Brazilian Participation Group, the Carnegie Institution for Science, 
Carnegie Mellon University, the Chilean Participation Group, the French Participation Group, Harvard-Smithsonian Center for Astrophysics, 
Instituto de Astrof\'isica de Canarias, The Johns Hopkins University, 
Kavli Institute for the Physics and Mathematics of the Universe (IPMU) / 
University of Tokyo, Lawrence Berkeley National Laboratory, 
Leibniz Institut f\"ur Astrophysik Potsdam (AIP), 
Max-Planck-Institut f\"ur Astronomie (MPIA Heidelberg), 
Max-Planck-Institut f\"ur Astrophysik (MPA Garching), 
Max-Planck-Institut f\"ur Extraterrestrische Physik (MPE), 
National Astronomical Observatories of China, New Mexico State University, 
New York University, University of Notre Dame, 
Observat\'ario Nacional / MCTI, The Ohio State University, 
Pennsylvania State University, Shanghai Astronomical Observatory, 
United Kingdom Participation Group,
Universidad Nacional Aut\'onoma de M\'exico, University of Arizona, 
University of Colorado Boulder, University of Oxford, University of Portsmouth, 
University of Utah, University of Virginia, University of Washington, University of Wisconsin, 
Vanderbilt University, and Yale University.
This work has made use of data from the European Space Agency (ESA)
mission {\it Gaia} (\url{https://www.cosmos.esa.int/gaia}), processed by
the {\it Gaia} Data Processing and Analysis Consortium (DPAC,
\url{https://www.cosmos.esa.int/web/gaia/dpac/consortium}). Funding
for the DPAC has been provided by national institutions, in particular
the institutions participating in the {\it Gaia} Multilateral Agreement.

\appendix
\section{Sources with [FeII] emission features}

\begin{figure*}
\begin{center}
\plottwo{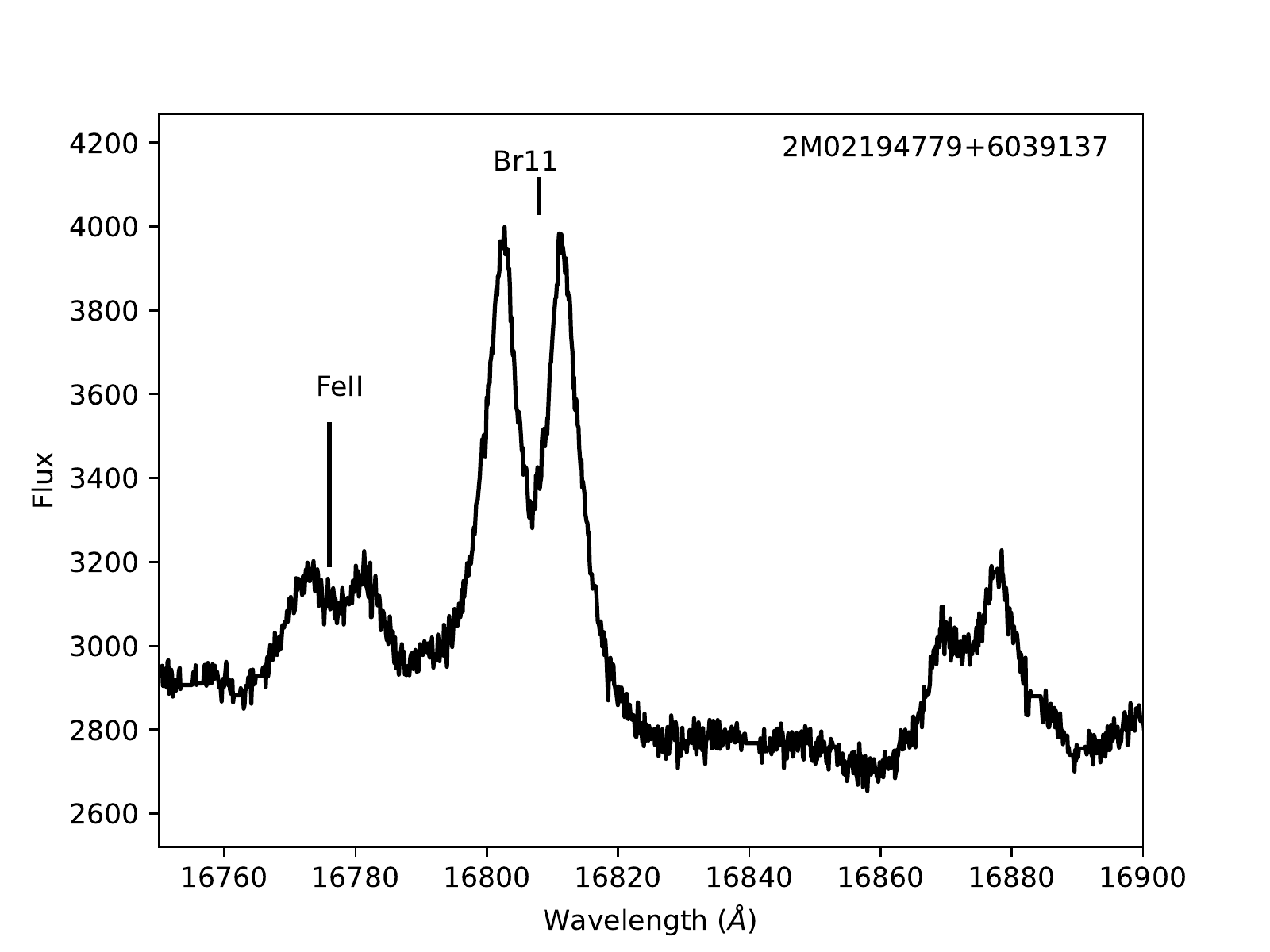}{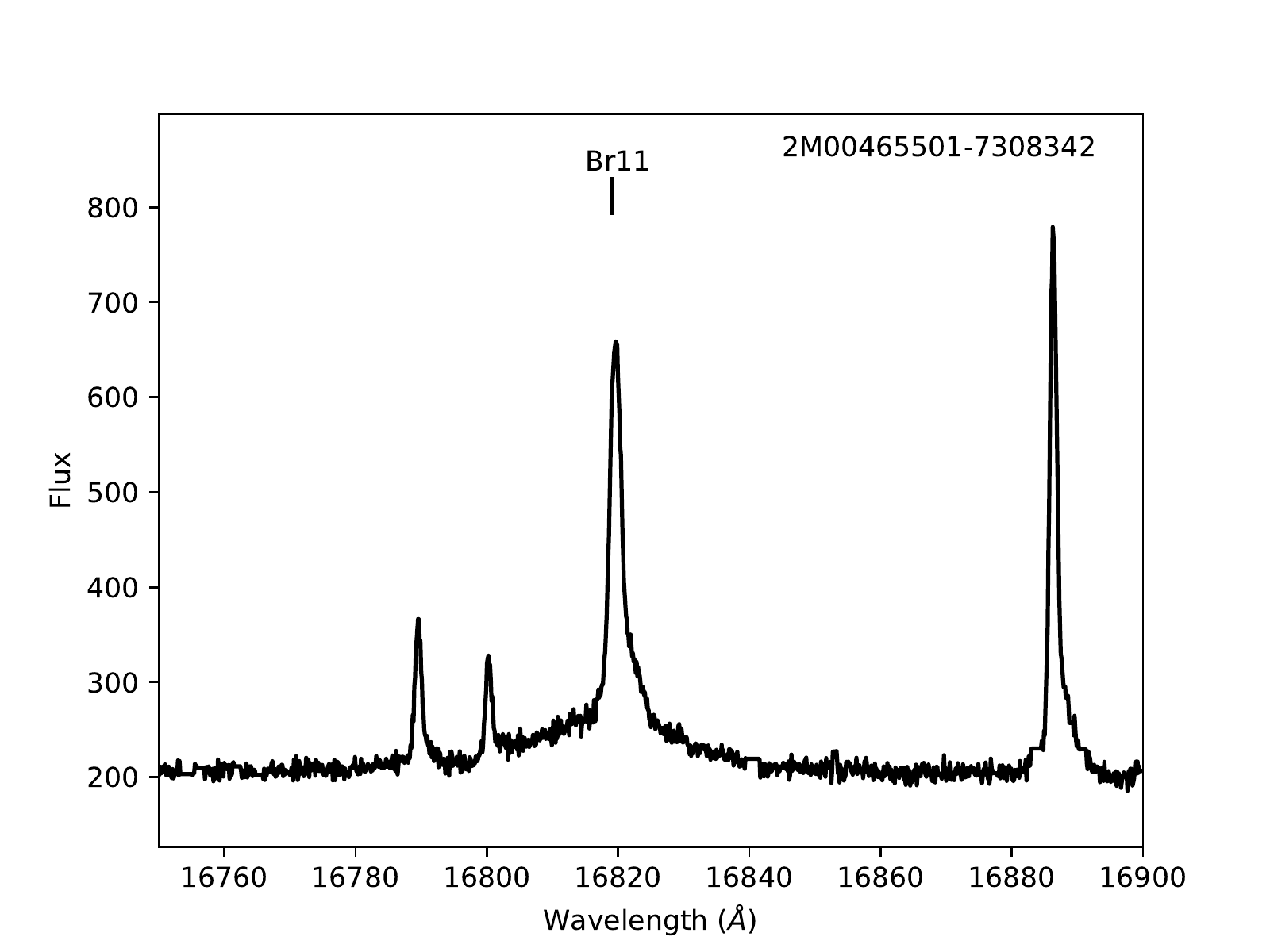}
\plottwo{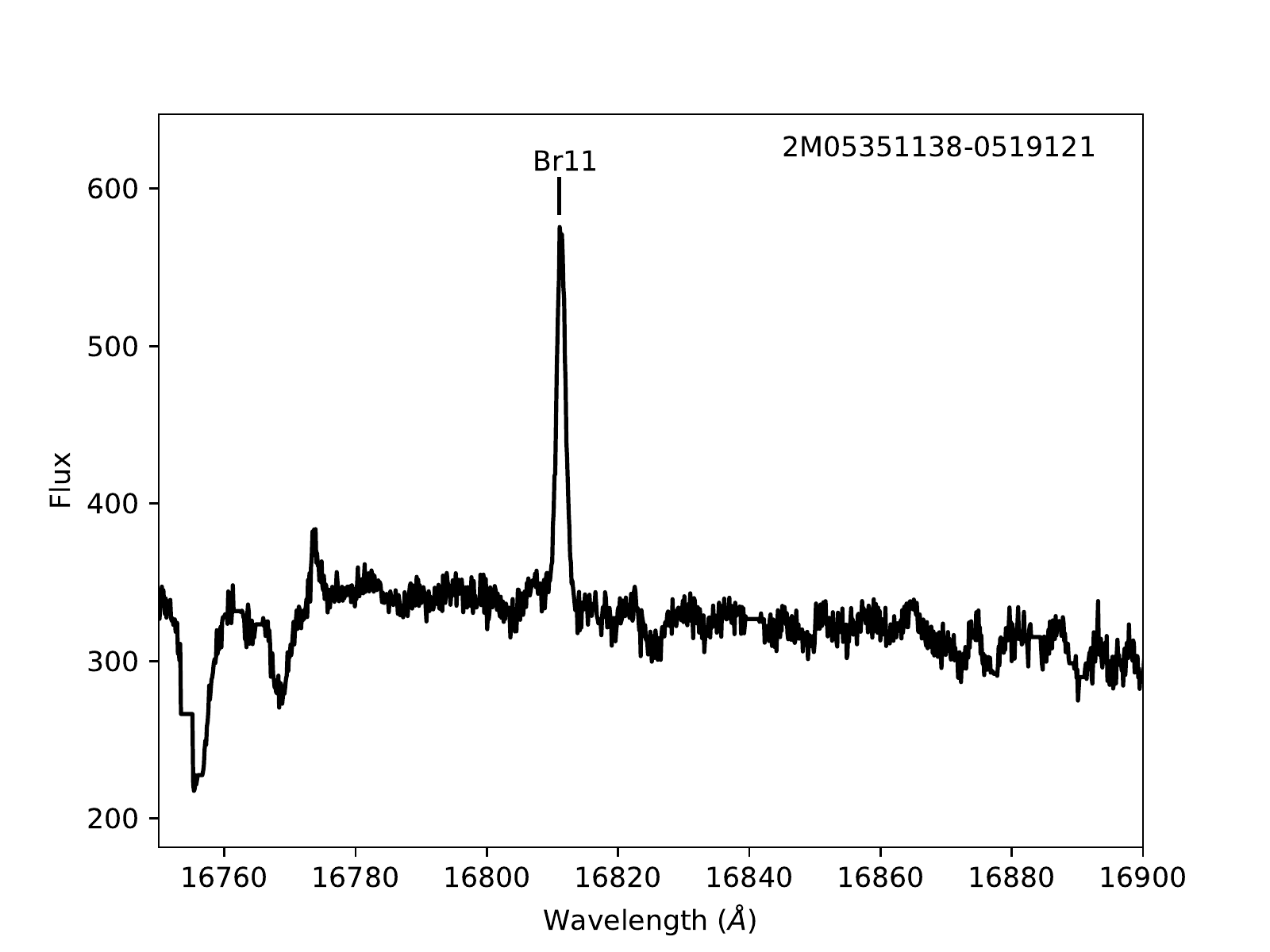}{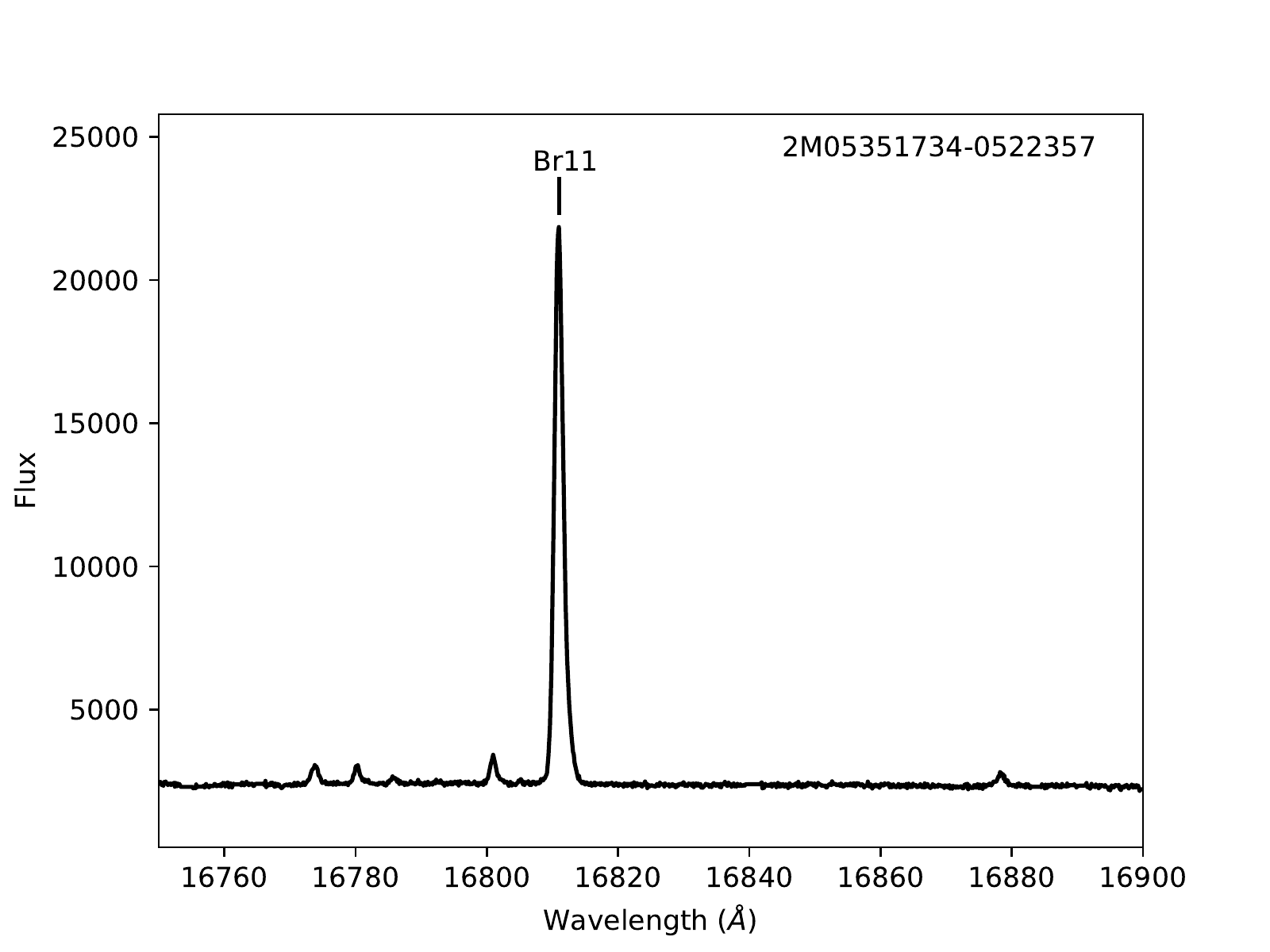}

\caption{\textit{(top left panel)}: Example of a source with (double-peaked) Br 11 emission, and a neighboring 16781 line with an identical velocity profile. These sources are coded with an FeII flag = 1 in the machine readable table of our sample.  \textit{(top right panel)}: Example of a source with strong Br 11 emission, and two lines near the wavelength expected for the neighboring [FeII] line. These sources are coded with an FeII flag = 2 in the machine readable table of our sample. \textit{(Bottom panels)}: Examples of two sources with a prominent emission feature at 16775: the source on the left shows this line near two photospheric features, while the source on the right shows two additional emission lines in between the 16775 line and Br11.  Both of these patterns which include the 16775 feature are coded with an FeII flag = 3 in the machine readable table of our sample. \label{fig:FeII_emitters}}

\end{center}
\end{figure*}

In the course of this analysis, we visually identified a number of sources with prominent [FeII] emission lines at 16790\AA~ (i.e., just blueward of the main Br11 emission line, Figure \ref{fig:FeII_emitters}), as well as sources with emission from unidentified lines at 16782 and 16776. The 16791 [FeII] line is more often seen in Be stars than YSOs, and a number of these [FeII] emission sources were previously identified in the analysis by \citet{chojnowski2015}. Given the expansion of the APOGEE dataset since the \citet{chojnowski2015} analysis, however, for completeness we list all sources for which we observe [FeII] emission lines in the 'FeII' column of the machine readable table of this sample. \\

Sources with a prominent 16781\AA~ emission feature with the same velocity profile as the neighboring Br11 line are listed in the table with a '1' in the FeII column.  There are 63 sources with this emission line, of which 38 are found on young cluster plates; an example of this type is shown in the upper left panel of Fig. \ref{fig:FeII_emitters}.

An additional 17 sources (but only 5 on young cluster plates) appear to have multiple, velocity separated 16792 Fe II lines, potentially indicating the presence of nebular emission as well as a jet. An example of this type of spectrum, which is coded as '2' in the machine readable table, is shown in the upper right corner of Fig. \ref{fig:FeII_emitters}.

Finally, another 37 sources (of which 36 are found on young cluster plates) show an emission feature near 16775.  Some of these sources include additional emission lines between the 16775 line and Br11, while others feature just relatively weak Br11 and 16775 lines.  Examples of these types of spectra, which are coded as '3' in the machine readable table, are shown in the bottom row of Fig. \ref{fig:FeII_emitters}.

\startlongtable
\begin{deluxetable*}{ccl}
\tablecaption{Line Measurements for APOGEE DR17 Br11 Emitters \label{tab:dr17_data}}
\tabletypesize{\scriptsize}
\tablewidth{\linewidth}
\tablehead{
 \colhead{Column} &
 \colhead{Unit} &
 \colhead{Description}
 }
\startdata
2MASS ID & & 2MASS source ID \\
Plate & & SDSS Plate ID \\
MJD &  & MJD of SDSS observation \\
Fiber & & SDSS Fiber ID \\
RA & deg & Right Ascension in J2000 \\
Dec & deg & Declination in J2000 \\
Telescope & & SDSS Telescope ID \\
Location & & SDSS Location ID \\
Field & & SDSS Field ID \\
Br11 Auto EqW & Angstroms & Scripted Equivalent Width Measurement of Br11 Emission Line \\
Br11 Auto Error & Angstroms & Uncertainty in Scripted Equivalent Width Measurement of Br11 Emission Line \\
Br12 Auto EqW & Angstroms & Scripted Equivalent Width Measurement of Br12 Emission Line \\
Br12 Auto Error & Angstroms & Uncertainty in Scripted Equivalent Width Measurement of Br12 Emission Line \\
Br13 Auto EqW & Angstroms & Scripted Equivalent Width Measurement of Br13 Emission Line \\
Br13 Auto Error & Angstroms & Uncertainty in Scripted Equivalent Width Measurement of Br13 Emission Line \\
Br14 Auto EqW & Angstroms & Scripted Equivalent Width Measurement of Br14 Emission Line \\
Br14 Auto Error & Angstroms & Uncertainty in Scripted Equivalent Width Measurement of Br14 Emission Line \\
Br15 Auto EqW & Angstroms & Scripted Equivalent Width Measurement of Br15 Emission Line \\
Br15 Auto Error & Angstroms & Uncertainty in Scripted Equivalent Width Measurement of Br15 Emission Line \\
Br16 Auto EqW & Angstroms & Scripted Equivalent Width Measurement of Br16 Emission Line \\
Br16 Auto Error & Angstroms & Uncertainty in Scripted Equivalent Width Measurement of Br16 Emission Line \\
Br17 Auto EqW & Angstroms & Scripted Equivalent Width Measurement of Br17 Emission Line \\
Br17 Auto Error & Angstroms & Uncertainty in Scripted Equivalent Width Measurement of Br17 Emission Line \\
Br18 Auto EqW & Angstroms & Scripted Equivalent Width Measurement of Br18 Emission Line \\
Br18 Auto Error & Angstroms & Uncertainty in Scripted Equivalent Width Measurement of Br18 Emission Line \\
Br19 Auto EqW & Angstroms & Scripted Equivalent Width Measurement of Br19 Emission Line \\
Br19 Auto Error & Angstroms & Uncertainty in Scripted Equivalent Width Measurement of Br19 Emission Line \\
Br20 Auto EqW & Angstroms & Scripted Equivalent Width Measurement of Br20 Emission Line \\
Br20 Auto Error & Angstroms & Uncertainty in Scripted Equivalent Width Measurement of Br20 Emission Line \\
YC Object & & Flag indicating a source is in the Young Cluster sample (=1 if so, 0 if not) \\
YC Density & n$_e$ cm$^{-3}$ & Electron Density of Best Fit Model to Brackett Decrement\\  & & (for sources on Young Cluster plates) \\
YC Max. Dens. & n$_e$ cm$^{-3}$ & Maximum Electron Density of models within XX of minimum chi square\\
& & (for sources on Young Cluster plates) \\
YC Min. Dens. & n$_e$ cm$^{-3}$ & Minimum Electron Density of models within XX of minimum chi square\\
& & (for sources on Young Cluster plates) \\
YC Dens. Std. Dev. & n$_e$ cm$^{-3}$ & Standard Deviation of Electron Densities for models within XX of minimum chi square\\
 & & (for sources on Young Cluster plates) \\
YC Temp. & deg (K) & Gas Temperature of Best Fit Model to Brackett Decrement\\
& & (for sources on Young Cluster plates) \\
YC Max. Temp. & deg (K) & Maximum Gas Temperature of all models within XX of minimum chi square\\
& & (for sources on Young Cluster plates) \\
YC Min. Temp. & deg (K) & Minimum Gas Temperature of all models within XX of minimum chi square\\
& & (for sources on Young Cluster plates) \\
YC Temp. Std. Dev. & deg (K) & Standard Deviation of Gas Temperature for all models within XX of minimum chi square \\
& & (for sources on Young Cluster plates) \\
YC Chi & & Chi Square of Best Fit Model to Brackett Decrement\\
  & & (for sources on Young Cluster plates) \\
YC Chi Dev. & & Standard Deviation of Chi Square values for fits within XXX of minimum chi square\\
  & & (for sources on Young Cluster plates) \\
slope &  & ratio of continuum fluxes near the Br 11 and Br 20 lines, presented as Br11$_{cont}$ \ Br20$_{cont}$\\
& & (values greater than 1 represent sources with SEDs rising to long wavelengths) \\
FeII & & FeII emission flag; 1 if one FeII line detected; 2 if two velocity separated FeII lines detected; 0 otherwise \\
NN Detected & & Neural Net Classification for Br Line Detection (1=line detected; 0=no line detected)\\
NN Double & & Neural Net Classification of Br Line Profiles (1=double peaked profile; 0=single peak profile) \\
NN Nebular & & Neural Net Classification for Br Line Width (1=narrow/ISM width; 0=Doppler broadened profile)\\
NN Good & & Neural Net Classification of Br Line Quality (1=Good; 0=Bad) \\
NN Center Vel. & km s$^{-1}$ & Neural Net Estimate of Br Line Center Velocities\\
NN Vel. Scatter & km s$^{-1}$ & Neural Net Estimate of Uncertainty in Br Line Center Velocities\\
NN Vel. Width & km s$^{-1}$ & Neural Net Estimate of Velocity Width of Br Line Profiles \\
NN Vel. Width Scatter & km s$^{-1}$ & Neural Net Estimate of Uncertainty in Br Line Velocity Width\\
NN Maxline &  & Highest order Brackett line detected by the Neural Net analysis\\
Br11 NN EqW & Angstroms & Neural Net Equivalent Width Measurement of Br11 Emission Line \\
Br11 NN Error & Angstroms & Uncertainty in Neural Net Equivalent Width Measurement of Br11 Emission Line \\
Br12 NN EqW & Angstroms & Neural Net Equivalent Width Measurement of Br12 Emission Line \\
Br12 NN Error & Angstroms & Uncertainty in Neural Net Equivalent Width Measurement of Br12 Emission Line \\
Br13 NN EqW & Angstroms & Neural Net Equivalent Width Measurement of Br13 Emission Line \\
Br13 NN Error & Angstroms & Uncertainty in Neural Net Equivalent Width Measurement of Br13 Emission Line \\
Br14 NN EqW & Angstroms & Neural Net Equivalent Width Measurement of Br14 Emission Line \\
Br14 NN Error & Angstroms & Uncertainty in Neural Net Equivalent Width Measurement of Br14 Emission Line \\
Br15 NN EqW & Angstroms & Neural Net Equivalent Width Measurement of Br15 Emission Line \\
Br15 NN Error & Angstroms & Uncertainty in Neural Net Equivalent Width Measurement of Br15 Emission Line \\
Br16 NN EqW & Angstroms & Neural Net Equivalent Width Measurement of Br16 Emission Line \\
Br16 NN Error & Angstroms & Uncertainty in Neural Net Equivalent Width Measurement of Br16 Emission Line \\
Br17 NN EqW & Angstroms & Neural Net Equivalent Width Measurement of Br17 Emission Line \\
Br17 NN Error & Angstroms & Uncertainty in Neural Net Equivalent Width Measurement of Br17 Emission Line \\
Br18 NN EqW & Angstroms & Neural Net Equivalent Width Measurement of Br18 Emission Line \\
Br18 NN Error & Angstroms & Uncertainty in Neural Net Equivalent Width Measurement of Br18 Emission Line \\
Br19 NN EqW & Angstroms & Neural Net Equivalent Width Measurement of Br19 Emission Line \\
Br19 NN Error & Angstroms & Uncertainty in Neural Net Equivalent Width Measurement of Br19 Emission Line \\
Br20 NN EqW & Angstroms & Neural Net Equivalent Width Measurement of Br20 Emission Line \\
Br20 NN Error & Angstroms & Uncertainty in Neural Net Equivalent Width Measurement of Br20 Emission Line \\
\enddata
\end{deluxetable*}

\bibliography{main.bbl}{}
\bibliographystyle{aasjournal}

\end{document}